\documentclass[reprint,aps]{revtex4-1}
\usepackage{epsfig,graphicx,  setspace}
\usepackage{amsthm, amsmath, amssymb}
\usepackage{txfonts}
\graphicspath{{graphics/}}
\newcommand{\N}{{\rm I\!N}}

\newcommand{\vsp}{\vspace*{3mm}}
\newcommand{\be}{\begin{equation}}
\newcommand{\ee}{\end{equation}}
\newcommand{\bd}{\begin{displaymath}}
\newcommand{\ed}{\end{displaymath}}

\newcommand{\bra}{\langle}
\newcommand{\ket}{\rangle}

\newcommand{\order}{{\cal O}}

\newcommand{\bnull}{\mbox{\boldmath $0$}}
\newcommand{\bc}{\mbox{\boldmath $c$}}

\newcommand{\bk}{\mbox{\boldmath $k$}}

\newcommand{\bx}{\mbox{\boldmath $x$}}

\begin{document}

\title{Unbiased degree-preserving randomisation of directed binary networks}
\author{ES Roberts$^{\dag\ddag}$ and ACC Coolen$^{\dag\ddag\S}$} 

\affiliation{
${\dag}~$Department of Mathematics, King's College London, The Strand,
London WC2R 2LS, United Kingdom}
\affiliation{
${\ddag~}$Randall Division of Cell and Molecular Biophysics, King's College London, New
Hunts House, London SE1 1UL, United Kingdom}
\affiliation{$\S~$
London Institute for Mathematical Sciences, 35a South St, Mayfair, London W1K 2XF, United Kingdom}

\begin{abstract}{ 
Randomising networks using a naive `accept-all' edge-swap algorithm is generally biased. 
Building on recent results for nondirected graphs, we construct an ergodic detailed balance Markov chain with non-trivial acceptance probabilities for directed graphs, which converges  to a strictly uniform measure and is based on edge swaps that conserve all in- and out-degrees. 
The acceptance probabilities can also be generalized  to define Markov chains that target any alternative desired measure on the space of directed graphs, in order to generate graphs with more sophisticated topological features. This is demonstrated by defining a process tailored to the production of directed graphs with specified degree-degree correlation functions. 
The theory is implemented numerically and tested on synthetic and biological network examples. 
} 
\end{abstract}
\maketitle
\section{Introduction}

\noindent
When seeking to assess the statistical relevance of observations made in real networks, there are three routes available. One could generate 
null-model networks for hypothesis testing from scratch, constrained by the values of observed parameters in the real network (e.g. using the Molloy-Reed stub joining method \cite{MolloyReed}, or the Barab\'{a}si-Albert preferential attachment model \cite{Barabasi-Albert}). Alternatively, one could generate null-model networks 
by randomising the original network, using dynamical rules that leave the values of relevant parameters invariant \cite{Rao96}. The final option is to use analytical methods to find ensemble averages for the observable of interest, see e.g. \cite{Squartini11_theory,Squartini11_application}.

The null-model approach is appealing in its conceptual simplicity. It effectively provides synthetic `data', which can be analysed in the same way as the real dataset. One can then learn which observed properties are particular to the real dataset, and which are common within the ensemble. 

Applications of network null-models are wide ranging and central to network science. \cite{shen-orr02} applies null models to identify over-represented `motifs' in the transcriptional regulation
network of {\em E. coli.} \cite{Power} discusses adapting the Watts-Strogartz method to generating random networks to model power grids. 
 \cite{Japan_interfirm} explores motifs found within an interfirm network. \cite{newman_social} uses network null-models to study social networks. \cite{Maslov02} compares topological properties of interaction and transcription regulatory networks in yeast with randomised `null model' networks and postulated that links between highly connected proteins are suppressed in protein interaction networks. \cite{Ecology} discusses the challenges of specifying a suitable matrix null model in the field of ecology. 

It is crucial that the synthetic networks generated as null models are representative of the underlying ensembles. Any inadvertent bias in the network generation process may invalidate the hypothesis test. 
It is therefore worrying that the two most popular methods to randomise or generate null networks are in fact known to be biased. The common implementation of the stub-joining method, where invalid edges are rejected but the process subsequently continues (as opposed to starting from the beginning of the whole process), is known to be biased \cite{King04,Klein-Hennig,citeulike:9790164}; in fact, even if upon invalid edge rejection the stub-joining process is restarted, it is not clear whether the graphs produced would be unbiased (we are not aware of any published proof). Similarly, the conventional `accept-all' edge swap process, see e.g. \cite{Itz2003}, is also known to be biased \cite{Coolen09}: graphs on which many swaps can be executed are generated more often. 
The effects of these biases may in the past not always have been serious \cite{Milo04}, but using biased algorithms for producing null models is fundamentally unsound, and unacceptable when there are rigorous unbiased alternatives  \cite{Coolen09}.

In this paper we build on the work of \cite{Coolen09} and \cite{Rao96} and define a Markov Chain Monte Carlo process, based on ergodic in- and out- degree preserving edge-swap moves that act on {\em directed} networks. We first calculate correct move acceptance probabilities for the process to sample the space of all allowed directed graphs uniformly. We then extend the theory in order for the process to evolve to any desired measure on the space of directed graphs. Attention is paid to adapting our results for efficient numerical implementation. We also identify under which  circumstances the error made by doing `accept all' edge swaps is immaterial. We apply our theory to real and synthetic networks. 

\section{An ergodic and unbiased randomisation process which preserves in- and out-degrees}

\subsection{Edge swap moves}
\label{section:edgeswap}

\begin{figure}[t]
\unitlength=0.20mm
\begin{picture}(300,170)
\put(0,40){\includegraphics[width=290\unitlength]{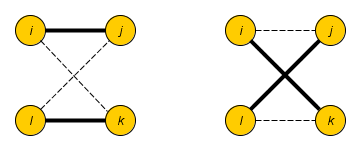}}
\put(40,20){before} \put(215,20){after}
\end{picture}
  \vspace*{-3mm}
  \caption{The undirected edge swap. This is the canonical choice for the elementary moves of an ergodic degree-preserving randomisation process on undirected networks. }
  \label{fig:undirected_swap}
\end{figure}

\noindent
The canonical moves for degree-preserving randomisation of graphs are the so-called `edge swaps', see e.g. \cite{ Seidel, Taylor,Coolen09}. The undirected version of the edge swap  is illustrated in figure \ref{fig:undirected_swap}; 
a generalisation to directed graphs is found in \cite{Rao96}. The authors of \cite{Rao96} define a move - which we will refer to as a \textit{square swap}  - starting from a set of four entries from the connectivity matrix $\bc\in\{0,1\}^{N^2}$ of a directed binary $N$-node graph, defined by node pairs $\{ (i_1,j_1),(i_1, j_2),(i_2, j_2),(i_2,j_1)  \}$ such that the corresponding entries  $\{ c_{i_1 j_1},c_{i_1, j_2},c_{i_2, j_2},c_{i_2,j_1}  \}$  are alternately $0$ and $1$, and not `structural' (i.e. they are allowed to vary). As for the undirected case, the elementary edge swap move is defined by swapping the $0$ and $1$ entries, i.e. $\{ c_{i_1 j_1},c_{i_1, j_2},c_{i_2, j_2},c_{i_2,j_1}  \}\to \{ 1\!-\!c_{i_1 j_1},1\!-\!c_{i_1, j_2},1\!-\!c_{i_2, j_2},1\!-\!c_{i_2,j_1}  \}$. The authors of \cite{Rao96} prove that, if self interactions are permitted,   repeated application of such moves can transform any binary matrix $\bc_A$ to any other binary matrix $\bc_B$ with the same  in- and out- degree distributions. 
\begin{figure}[t]
\unitlength=0.20mm
\hspace*{-3mm}
\begin{picture}(450,320)
\put(85,55){\includegraphics[width=280\unitlength]{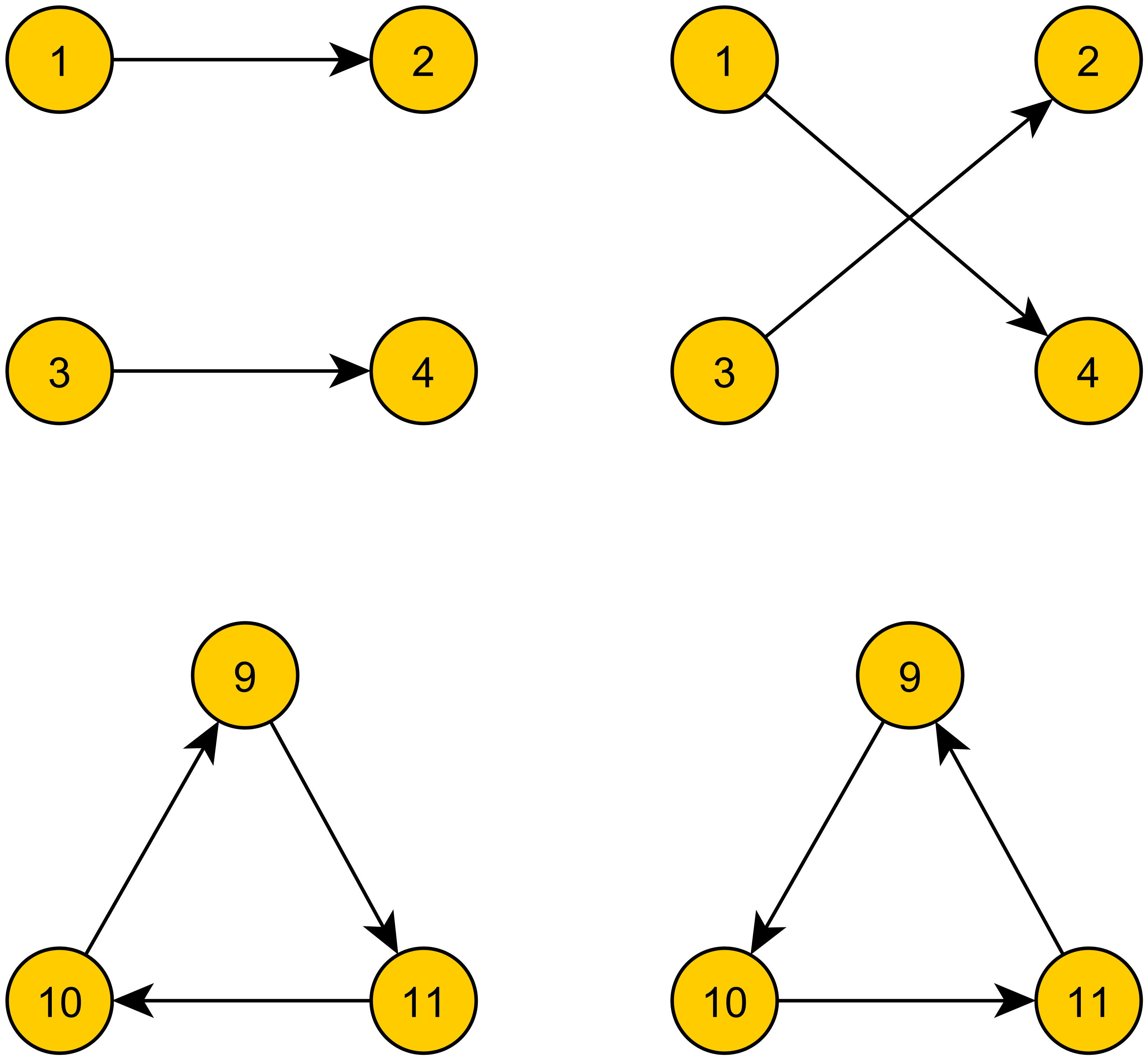}}
\put(120,20){before} \put(290,20){after}
\end{picture}
  \vspace*{-7mm}
  \caption{The \textit{square swap} (top) and \textit{triangle swap} (bottom). In combination these two represent the canonical choice for the elementary moves of an ergodic degree-preserving randomisation process on directed networks without self-interactions. }
\label{fig:Rao_moves}
\end{figure}
However, 
if we require in addition that the diagonal elements of all $\bc$ are 0 (i.e.we  forbid self-interactions), then the edge swap defined above is no longer sufficient to ensure ergodicity. To remedy this problem the authors of \cite{Rao96} introduce a further move, which we will call a \textit{triangle swap}. This move also gives us the simplest demonstration of two valid configurations that cannot be connected by square-type swaps. The \textit{square swap} and the \textit{triangle swap} are illustrated in  figure \ref{fig:Rao_moves}; in combination these two moves  allow us to transform between any two directed binary matrices which have the same in- and out-degrees, even if self-interactions are forbidden \cite{Rao96}. 

A stochastic process defined as accepting all randomly selected moves from the above set is ergodic but biased. This was already observed in \cite{Rao96}, where the authors proposed a `Switch \& Hold' algorithm, which involves the number of states accessible in one move from a configuration (its mobility), and the maximum possible number of states accessible in one move  from any network in the ensemble (the degrees of a hyper-graph, in the language of later publications). 
In \cite{Coolen09} the problem was studied for undirected graphs; it was shown how move acceptance probabilities should be defined to guarantee stochastic evolution by edge swapping to any desired measure on the space of nondirected graphs. The analysis in \cite{Coolen09} is quite general, and briefly reproduced in section \ref{sec:theory} below. Here we will adapt their calculations to directed graphs and include the new moves defined by \cite{Rao96}. This will result in a Markovian process based on edge swapping that will equilibrate to any desired measure on the space of {\em directed} graphs.

\subsection{Outline of the general theory}
\label{sec:theory}

\noindent
This section briefly summarizes results of \cite{Coolen09} which will be used in the next section. We define an adjacency matrix $\bc = \{ c_{ij}\}$, where $c_{ij}=1$ if and only if there is a directed link from node $j$ to node $i$. We denote the set of all such graphs as $C$. The aim is to define and study constrained Markov chains for the evolution of $\bc$ in some subspace $\Omega \in C$. This is a discrete time stochastic process, where the probability $p_{t}(\bc)$  of observing a graph $\bc$ at time $t$ evolves according to

\begin{eqnarray}
 \forall  \bc \in \Omega ~~~:~~~ p_{t+1}(\bc) = \sum_{\bc^{\prime} \in \Omega} W(\bc | \bc^\prime) p_t (\bc^\prime)
 \end{eqnarray}
where $t \in \N$ and $W(\bc | \bc^\prime) $ is a transition probability. We require the process to have three additional properties: 
\begin{enumerate}
\item{Each elementary move $F$ can only act on a subset of all possible graphs. }
\item{The process converges to the invariant measure $$p_\infty(\bc) = Z^{-1} e ^{-H(\bc)}$$}
\item{Each move $F$ has a unique inverse, which acts on the same subset of states as $F$ itself. }
\end{enumerate}
We exclude the identity move from the set $\Phi$ of all moves, and we define an indicator function $I_F(\bc)$ where $I_F(\bc)=1$  iff the move $\bc \rightarrow F \bc$ is allowed.
The transition probabilities are constructed to obey detailed balance
\begin{eqnarray}
\forall \bc, \bc^\prime \in \Omega ~~:~~ W(\bc | \bc^\prime) p_\infty(\bc^\prime) = W(\bc^\prime | \bc)p_\infty(\bc) 
\label{eq:DB}
\end{eqnarray}
At each step a candidate move $F \in \Phi$ is drawn with probability $q(F|\bc^\prime)$, where $\bc^\prime$ is the current state. The move is accepted with some probability $A(F\bc^\prime|\bc^\prime)$.  In combination this leads to
\begin{eqnarray}
W(\bc|\bc^\prime) &=& \sum_{F \in \Phi} q(F|\bc^\prime) \left[\delta_{\bc, F\bc^\prime}A(F\bc^\prime|\bc^\prime)
+ \delta_{\bc, \bc^\prime}\left[1\!-\!A(F\bc^\prime|\bc^\prime)\right] \right]
\nonumber
\\[-2mm]&&
\end{eqnarray}
Insertion into (\ref{eq:DB}) then leads to the following conditions which must be satisfied by $A(F\bc^\prime|\bc^\prime)$ and $q(F|\bc^\prime)$:
\begin{eqnarray}
\label{eq:detailed_balance}
(\forall \bc \in \Omega)(\forall F \in \Phi): & \\[1mm] \nonumber
\! \!\!\!\!\! \! \!\!\!\!\! & \! \!\!\!\!\! \! \!\!\!\!\! \!\!\!\!\! \! \!\!\!\!\!\!\!\!\!\! \! \!\!\!\!\!
q(F|\bc) A(F\bc|\bc)e^{-H(\bc)}= q(F^{-1}|F\bc) A(\bc|F\bc)e^{-H(F\bc)}
\end{eqnarray}
We define the mobility $n(\bc)$ to be the number of moves which can act on each state: $n(\bc) = \sum_{F \in \Phi} I_F(\bc)$. If the candidate moves are drawn randomly with equal probabilities from the set of all moves allowed to act, we find 
(\ref{eq:detailed_balance}) reducing to
\begin{eqnarray}
 A(\bc|\bc^\prime) 
= \frac{n(\bc^\prime) e^{-\frac{1}{2}[H(\bc)-H(\bc^\prime)]}}{n(\bc^\prime) e^{-\frac{1}{2}[H(\bc)-H(\bc^\prime)]}+n(\bc) e^{\frac{1}{2}[H(\bc)-H(\bc^\prime)]}}
\label{eq:acceptance}
\end{eqnarray}
If we make the simplest choice  $H(\bc)= const$, the above process will asymptotically sample all graphs with the imposed degree sequence uniformly. To sample this constrained space of graphs with alternative nontrivial probabilities $p_\infty(\bc)$ we would choose $H(\bc)=-\log p_\infty(\bc)+const$.

Equation \ref{eq:detailed_balance} also shows what would happen if we were to sample with $A(\bc|\bc^\prime)  \equiv 1$ for all $(\bc,\bc^\prime)$, i.e. for  `accept all' edge swapping:  the detailed balance condition would give
\begin{eqnarray}
(\forall \bc \in \Omega)(\forall F \in \Phi): & ~~~~&
e^{-H(\bc)}n(\bc)= e^{-H(F\bc)}n(F\bc)
\end{eqnarray}
For this to be satisfied we require both sides of the expression to evaluate to a constant. Hence $e^{-H(\bc)} \propto n(\bc)$, so the naive process will converge to the non-uniform measure 
\begin{eqnarray}
p_\infty(\bc) = Z^{-1} n(\bc)
\end{eqnarray}
This is the undesirable bias of `accept-all' edge-swapping.  It has a clear intuitive explanation. The mobility $n(\bc)$ is the number of allowed moves on network $\bc$,  which is equal to the number of inverse moves through which $\bc$ can be reached in one step from another member of the ensemble. The likelihood of seeing a network $\bc$ upon equilibration of the process is proportional to the number of entry points that $\bc$  offers the process. 

\subsection{Calculation of the mobilities for directed networks}

\noindent
Since the two types of moves required for ergodic evolution of directed graphs, viz. the square swap and the triangle swap, are clearly distinct, the mobility of a graph $\bc$ is given by $n(\bc)=n_{\square}(\bc)+n_{\triangle}(\bc)$, where  $n_{\square}(\bc)$ and $n_{\triangle}(\bc)$ count the number of possible moves of each type that can be executed on $\bc$. 

To find $n_{\square}(\bc)$ we need to calculate how many distinct link-alternating cycles of length 4 can be chosen in graph $\bc$. We exclude  self-interactions, so our cycles must involve 4 distinct nodes. The total number of such moves can be written as
\begin{eqnarray}
n_{\square}(\bc)= \frac{1}{2}\sum_{ijk\ell}\bar{\delta}_{jk}\bar{\delta}_{\ell i}\bar{\delta}_{ik}\bar{\delta}_{j\ell} c_{ij}c_{k\ell}\bar{c}_{kj}\bar{c}_{i\ell}
\label{eq:nsqr}
\end{eqnarray}
where the pre-factor compensates for the symmetry, and where we used the short-hands $\bar{c}_{kj}= 1-{c}_{kj}$ and $\bar{\delta}_{jk} = 1- {\delta}_{jk}$. Expanding these shorthands in (\ref{eq:nsqr}) gives after some further bookkeeping of terms, and with $(\bc^\dag)_{ij}=c_{ji}$:
\begin{eqnarray}
n_{\square}(\bc)&=&
\frac{1}{2}{\rm Tr}(\bc \bc^\dag\bc\bc^\dag)
-\sum_{ij} k^{\rm in}_i c_{ij} k^{\rm out}_{j}
+{\rm Tr}(\bc\bc^\dag\bc)+\frac{1}{2}{\rm Tr}(\bc^2)
\nonumber
\\
&&
\hspace*{10mm}+\frac{1}{2}N^2\bra k\ket^2
-\sum_{j}k^{\rm in}_{j}k^{\rm out}_{j}
\label{eq:square_mobility_term}
\end{eqnarray}
with  $\bra k\ket=N^{-1}\sum_{i}k^{\rm in}_{i}=N^{-1}\sum_{i}k^{\rm out}_{i}$. 
We next repeat the calculation for the case of the \textit{triangle swap}. For easier manipulations, we introduce a new matrix $\mathbf{c^{\updownarrow}}$ of double links, defined via $(\bc^{\updownarrow})_{ij}= c_{ij}c_{ji}$. We then find
\begin{eqnarray}
n_{\triangle}(\bc)&=& \frac{1}{3}\sum_{ijk}\bar{\delta}_{ij}\bar{\delta}_{jk}\bar{\delta}_{ki} c_{ij} c_{jk} c_{ki}\bar{c}_{ji}\bar{c}_{kj}\bar{c}_{ik}
\nonumber
\\
&=& \frac{1}{3}\Big\{
{\rm Tr}(\bc^3)-3{\rm Tr}(\bc^{\updownarrow}\bc^2)+3{\rm Tr}(\bc^{\updownarrow 2}\bc)+-{\rm Tr}(\bc^{\updownarrow 3})\Big\}
\nonumber
\\
&=& \frac{1}{3}{\rm Tr}\big((\bc-\bc^\updownarrow)^3\big)
\label{eq:triangle_mobility_term}
\end{eqnarray}
In combination, (\ref{eq:square_mobility_term}) and (\ref{eq:triangle_mobility_term}) give us an explicit and exact formula for the graph mobility 
$n(\bc)=n_{\square}(\bc)+n_{\triangle}(\bc)$, and hence via (\ref{eq:acceptance}) a fully exact MCMC process for generating random graphs with 
prescribed sequences and any desired probability measure in the standard form $Z^{-1}\exp[-H(\bc)]$. 
Since (\ref{eq:square_mobility_term},\ref{eq:triangle_mobility_term}) cannot be written in terms of the degree sequence only, neglecting the mobility  (as with accept-all edge swapping) would  always introduce a bias into the sampling process.

\section{Properties and impact of graph mobility}

\subsection{Bounds on the mobility}
\label{sec:bounds}

\noindent
We will now derive bounds on the sizes of the mobility terms. This may show for which types of networks the application of `accept all' edge swapping (which ignores the mobility terms) is most dangerous, and for which networks the unwanted bias may be small. 
We first observe that 
\begin{eqnarray*}
n_\triangle(\bc) 
 = \frac{1}{3} \sum_{ijk}c_{ij}(1 - c_{ji})c_{jk}(1 - c_{kj})c_{ki}(1 - c_{ik}) 
\leq \frac{1}{3} {\rm Tr}(\bc^3) 
\end{eqnarray*}
Hence, the mobility $n(\bc)=n_\square(\bc) +n_\triangle(\bc)$ obeys
\begin{eqnarray}
n(\bc)&\leq &
\frac{1}{2}{\rm Tr}(\bc \bc^\dag\bc\bc^\dag)
-\sum_{ij} k^{\rm in}_i c_{ij} k^{\rm out}_{j}
+{\rm Tr}(\bc\bc^\dag\bc)+\frac{1}{2}{\rm Tr}(\bc^2)
\nonumber
\\
&&
\hspace*{10mm}+\frac{1}{2}N^2\bra k\ket^2
-\sum_{j}k^{\rm in}_{j}k^{\rm out}_{j}
+ \frac{1}{3} {\rm Tr}(\bc^3) 
\label{eq:n_upper1}
\end{eqnarray}
We find  upper bounds for most of the terms above by applying the simple inequality 
$c_{ij}c_{kl} \leq \frac{1}{2}(c_{ij}+c_{kl})$, which gives e.g.
\begin{eqnarray}
\hspace*{-6mm}
{\rm Tr}(\bc\bc^\dag \bc) 
		\leq 
		\frac{N}{2} \left[ \langle k^{\rm in ~2} \rangle + \langle k^{\rm out ~2} \rangle \right]  &~~~~~~& 
		{\rm Tr}(\bc^2) \leq  N \bra k\ket 
		\label{eq:bounds1}
\\
\hspace*{-6mm}
{\rm Tr}(\bc \bc^\dag \bc \bc^\dag) \leq \sum_{ij} k_i^{\rm in}c_{ij}k_i^{\rm out}~~~~~~~~
&~~~~~~& 
{\rm Tr}(\bc^3) \leq  N \bra k^{\rm in} k^{\rm out} \ket 
\label{eq:bounds2}
\end{eqnarray}
An upper bound on $\sum_{ij} k_i^{\rm in}c_{ij}k_j^{\rm out}$ follows from the observation that if $c_{ij} = 1$ then certainly $k_i^{\rm in} \geq 1$ and $k_j^{\rm out} \geq 1$.
Hence
\begin{eqnarray}
 \sum_{ij} k_i^{\rm in}c_{ij}k_j^{\rm out}& \geq&  \frac{1}{2}\sum_{ij} [c_{ij}k_j^{\rm out}+k_i^{\rm in}c_{ij}]
 \nonumber\\
 &=&
 \frac{1}{2}N [\langle k^{\rm in~2} \rangle +  \langle k^{\rm out~2} \rangle ]
 \label{eq:bounds3}
\end{eqnarray}
Combining (\ref{eq:bounds1},\ref{eq:bounds2},\ref{eq:bounds3}) with (\ref{eq:n_upper1}) then gives 
\begin{eqnarray}
n(\bc)&\leq &
\frac{N}{2}\Big[
N\bra k\ket^2+ \bra k\ket +
\frac{1}{2} [ \langle k^{\rm in ~2} \rangle \!+\! \langle k^{\rm out ~2} \rangle ]
- \frac{4}{3}  \bra k^{\rm in} k^{\rm out} \ket \Big]
\nonumber
\\[-1mm]
&&
\label{eq:final_upper_bound}
\end{eqnarray}
Next we calculate a lower bound for $n(\bc)$. 
We use simple identities such as
\begin{eqnarray}
{\rm Tr}(\bc^2) \geq 0 ~~~~~~~~~~~
n_\triangle(\bc) \geq 0 ~~~~~~~~~~~
{\rm Tr}(\bc\bc^\dag \bc) \geq 0 ~~~~~~~~~~~
\end{eqnarray}
and 
\begin{eqnarray}
{\rm Tr}(\bc \bc^\dag \bc \bc^\dag)
&\geq&\frac{1}{2} \sum_{ijk\ell} c_{ji}c_{jk}c_{\ell k}c_{\ell i} (\delta_{j\ell} +\delta_{ik})\nonumber
\\
&=& N[\bra k^{\rm in~2}\ket+\bra k^{\rm out~2}\ket]
\end{eqnarray}
We now find 
\begin{eqnarray}
n(\bc)&\geq &
\frac{1}{2}N[\bra k^{\rm in~2}\ket+\bra k^{\rm out~2}\ket]
+\frac{1}{2}N^2\bra k\ket^2
-\sum_{j}k^{\rm in}_{j}k^{\rm out}_{j}
\nonumber
\\
&&\hspace*{10mm}-\sum_{ij} k^{\rm in}_i c_{ij} k^{\rm out}_{j}
\end{eqnarray}
We finally need an upper bound for $\sum_{ij} k^{\rm in}_i c_{ij} k^{\rm out}_{j}$, which we write in terms of 
$k_{\rm max}^{\rm in}=\max_i k_i^{\rm in}$ and $k_{\rm max}^{\rm out}=\max_i k_i^{\rm out}$:
\begin{eqnarray}
\sum_{ij} k^{\rm in}_i c_{ij} k^{\rm out}_{j}&\leq & \frac{1}{2}\sum_{ij} [k^{\rm in}_{\rm max} c_{ij} k^{\rm out}_{j}+k^{\rm in}_i c_{ij} k^{\rm out}_{\rm max}]
\nonumber\\
&=&  \frac{1}{2}N\Big[ k^{\rm in}_{\rm max}\bra k^{\rm out~2}\ket 
+k^{\rm out}_{\rm max}\bra k^{\rm in~2}\ket
\Big]
\end{eqnarray}
We thus obtain our lower bound for the mobility:
\begin{eqnarray}
n(\bc)
&\geq &
\frac{N}{2}\left[N\bra k\ket^2
+\bra (k^{\rm in}\!-\! k^{\rm out})^2\ket
- k^{\rm in}_{\rm max}\bra k^{\rm out~2}\ket 
-k^{\rm out}_{\rm max}\bra k^{\rm in~2}\ket
\right]
\nonumber
\\[-1mm]
&& \label{eq:final_lower_bound}
\end{eqnarray}

\subsection{Identification of graph types most likely to be biased by `accept all' edge swapping}

\noindent
We know from (\ref{eq:acceptance}) that unbiased sampling of graphs, i.e. $p(\bc)= 1/|\Omega|$ for all $\bc\in\Omega$, requires using the following state-dependent acceptance probabilities in the edge swap process:
\begin{eqnarray}
 A(\bc|\bc^\prime) 
&=& [1+n(\bc)/n(\bc^\prime)]^{-1}
\label{eq:togetflat}
\end{eqnarray}
We now investigate under which conditions one will in large graphs effectively find $n(\bc)/n(\bc^\prime)\to 1$ for all $\bc,\bc^\prime\in\Omega$, so 
that the sampling bias would be immaterial. Let us define 
\begin{eqnarray}
\Delta n&=&\max_{\bc,\bc^\prime\in\Omega} |n(\bc)-n(\bc^\prime)|=\max_{\bc\in\Omega}n(\bc)-\min_{\bc\in\Omega}n(\bc)
\end{eqnarray}
Using the two bounds (\ref{eq:final_upper_bound},\ref{eq:final_lower_bound}) we immediately obtain
\begin{eqnarray}
\Delta n&\leq &
\frac{N}{2}\Big[
 \bra k\ket -
\frac{1}{2} [ \langle k^{\rm in ~2} \rangle \!+\! \langle k^{\rm out ~2} \rangle ]
+ \frac{2}{3}  \bra k^{\rm in} k^{\rm out} \ket 
\nonumber
\\
&&
\hspace*{15mm} + k^{\rm in}_{\rm max}\bra k^{\rm out~2}\ket 
+k^{\rm out}_{\rm max}\bra k^{\rm in~2}\ket
\Big]
\nonumber
\\
&=& 
\frac{N}{2}\Big[
 \bra k\ket
  -\frac{1}{6} [ \langle k^{\rm in ~2} \rangle \!+\! \langle k^{\rm out ~2} \rangle ]
-\frac{1}{3}\bra (k^{\rm in}- k^{\rm out})^2\ket
\nonumber
\\
&&\hspace*{15mm} 
+ k^{\rm in}_{\rm max}\bra k^{\rm out~2}\ket 
+k^{\rm out}_{\rm max}\bra k^{\rm in~2}\ket
\Big]
\nonumber
\\
&\leq & \frac{N}{2}\Big[
 \bra k\ket
+ k^{\rm in}_{\rm max}\bra k^{\rm out~2}\ket 
+k^{\rm out}_{\rm max}\bra k^{\rm in~2}\ket
\Big]
\end{eqnarray}
Clearly $1-\Delta n/n(\bc)\leq n(\bc^\prime)/n(\bc)\leq 1+\Delta n/n(\bc)$, so in view of (\ref{eq:togetflat}) 
we are interested in the ratio $\Delta n/n(\bc)$, for which we find
\begin{eqnarray}
\frac{\Delta n}{n(\bc)}&\leq &
\frac{
 \bra k\ket
+ k^{\rm in}_{\rm max}\bra k^{\rm out~2}\ket 
+k^{\rm out}_{\rm max}\bra k^{\rm in~2}\ket
}{
N\bra k\ket^2
- k^{\rm in}_{\rm max}\bra k^{\rm out~2}\ket 
-k^{\rm out}_{\rm max}\bra k^{\rm in~2}\ket
}
\end{eqnarray}
So  we can be confident that the impact of the graph mobility on the correct acceptance probabilities 
(\ref{eq:togetflat}) is  immaterial if
\begin{eqnarray}
 \frac{1}{\bra k\ket}+
 \frac{2}{\bra k\ket^2}\Big(k^{\rm in}_{\rm max}\bra k^{\rm out~2}\ket 
+k^{\rm out}_{\rm max}\bra k^{\rm in~2}\ket\Big)
\ll
N
\label{eq:final_condition}
\end{eqnarray}
We see from this that we can apply the `accept all' edge-swap process with confidence when we are working with a large network with a narrow degree distribution. 


\section{Mobilities of simple graph examples}

\noindent
In this section we confirm the validity of the mobility formulae (\ref{eq:square_mobility_term},\ref{eq:triangle_mobility_term}) for several simple 
examples of directed graphs.
\begin{enumerate}
\item {\em Two isolated bonds:}\\[1mm]
Here we have  $c_{12}=1$, $c_{34}=1$, and $c_{ij}=0$ for all $(i,j)\notin\{(1,2),(3,4)\}$. It is immediately clear that $\bc^\updownarrow=\bnull$, and 
\begin{eqnarray*}
&~&\sum_{ij} k_i^{\rm in} c_{ij} k_j^{\rm out} = 2,~~~
\sum_j k_j^{\rm out}k_j^{\rm in}=0,~~~\bra k\ket=\frac{2}{N}
\hspace*{-3mm}
\\
&~&
{\rm Tr}(\bc^2)={\rm Tr}(\bc^3)={\rm Tr}(\bc\bc^\dag \bc)=0,~~~{\rm Tr}(\bc \bc^\dag\bc\bc^\dag)=2
\hspace*{-3mm}
\end{eqnarray*}
Insertion into (\ref{eq:square_mobility_term},\ref{eq:triangle_mobility_term}) gives $n_{\square}(\bc)=1 $ and $n_{\triangle}(\bc)=0$. 
As we would expect: only one (square) move is permitted. 

\item {\em Isolated triangle:}
\\[1mm]
This example is defined by $c_{12}=c_{23}=c_{31}=1$, with $c_{ij}=0$ for all $(i,j)\notin\{(1,2),(2,3),(3,1)\}$. 
Again we have $\bc^\updownarrow=\bnull$, but now
\begin{eqnarray*}
&~&\sum_{ij} k_i^{\rm in} c_{ij} k_j^{\rm out} = 3,~~~
\sum_j k_j^{\rm out}k_j^{\rm in}=3,~~~\bra k\ket=\frac{3}{N}
\\
&~&{\rm Tr}(\bc^2)={\rm Tr}(\bc\bc^\dag \bc)=0,~~~{\rm Tr}(\bc^3)=3,~~~
{\rm Tr}(\bc \bc^\dag\bc\bc^\dag)=3
\hspace*{-10mm}
\end{eqnarray*}
This results in $n_{\square}(\bc)=0$ and   $n_{\triangle}(\bc)=1$. The only possible move is reversal of the directed triangle. 

\item {\em Complete (fully connected) graph:}
\\[1mm]
Here  $c_{ij}= 1 - \delta_{ij}$, and no edge swaps are possible. All nodes have $k_i^{\rm in}=k_i^{\rm out}= N-1$, and since $\bc^\updownarrow=\bc$ we know that $n_{\triangle}(\bc)=0$. This connectivity matrix, also featured in \cite{Coolen09}, has eigenvalues $\lambda= N-1$ (multiplicity 1) and $\lambda = -1$ (multiplicity $N-1$). Hence 
\begin{eqnarray*}
&~&\sum_{ij} k_i^{\rm in} c_{ij} k_j^{\rm out} = N(N\!-\!1)^3,~~~
\sum_j k_j^{\rm out}k_j^{\rm in}=N(N\!-\!1)^2\hspace*{-10mm}
\\
&~&{\rm Tr}(\bc^2)=\sum_i\lambda_i^2=N(N\!-\!1),\\
&~&{\rm Tr}(\bc\bc^\dag \bc)={\rm Tr}(\bc^3)=\sum_i\lambda_i^3=N(N\!-\!1)(N\!-\!2),\\
&~&{\rm Tr}(\bc \bc^\dag\bc\bc^\dag)={\rm Tr}(\bc^4)=\sum_i\lambda_i^4=(N\!-\!1)[(N-1)^3+1]
\hspace*{-10mm}
\end{eqnarray*}
Assembling the entire expression for the square mobility term (\ref{eq:square_mobility_term}) indeed gives  the correct value $n_{\square}(\bc)=0 $. 

\item {\em Directed spaning ring:}
\\[1mm]
This directed graph, defined by $c_{ij}= \delta_{i+1,j}$  modulo $N$,  gives a ring with a flow around it. We choose $N>2$.
Once more $\bc^\updownarrow=\bnull$, and we obtain for the relevant terms
\begin{eqnarray*}
&~&\sum_{ij} k_i^{\rm in} c_{ij} k_j^{\rm out} =\sum_j k_j^{\rm out}k_j^{\rm in}= N,~~~
\bra k\ket=1
\hspace*{-3mm}
\\
&~&
{\rm Tr}(\bc^2)={\rm Tr}(\bc^3)={\rm Tr}(\bc\bc^\dag \bc)=0,~~~{\rm Tr}(\bc \bc^\dag\bc\bc^\dag)=N
\hspace*{-3mm}
\end{eqnarray*}
 The final result, $n_{\square}(\bc)=  \frac{1}{2} N (N-3) $ and $n_{\triangle}(\bc)=0$,  is again what we would expect. As soon as one first bond to participate in an edge swap is picked (for which there are $N$ options),  there are $N-3$ possibilities for the second (since the already picked bond and its neighbours are forbidden). The factor 2 corrects for over-counting.

\item {\em Bidirectional spanning ring:}
\\[1mm]
Our final example is the nondirected version of the previous graph, viz. 
 $c_{ij} = \delta_{i, j-1} + \delta_{i, j+1}$ modulo $N$, with $N>2$. Since $\bc^\updownarrow=\bc$ we have $n_{\triangle}(\bc)=0$. 
Now 
\begin{eqnarray*}
&~&\sum_{ij} k_i^{\rm in} c_{ij} k_j^{\rm out}=8N,~~~ \sum_j k_j^{\rm out}k_j^{\rm in}= 4N,~~~
\bra k\ket=2
\hspace*{-3mm}
\\
&~&
{\rm Tr}(\bc^2)=2N,~~~{\rm Tr}(\bc^3)={\rm Tr}(\bc\bc^\dag \bc)=0
\\
&~&{\rm Tr}(\bc \bc^\dag\bc\bc^\dag)=6N
\hspace*{-3mm}
\end{eqnarray*}
We thereby find $n_{\square}(\bc)= 2N(N-4)$. 
 This is double the mobility evaluated in \cite{Coolen09}, since every move in the undirected version of the network corresponds to two possible moves in the directed version of the network. 
\end{enumerate}


\section{A randomisation process which preserves degrees 
and targets degree-degree correlations}

\noindent
So far we 
applied formula (\ref{eq:acceptance}) for the canonical acceptance probabilities for directed graph edge swapping to the problem of generating graphs with prescribed in- and out-degrees $(\bk^{\rm in},\bk^{\rm out})$ and a uniform measure. Here consider how to generate graphs which, in addition, display certain degree correlations. We first rewrite (\ref{eq:acceptance})  as
\begin{eqnarray}
 A(\bc|\bc^\prime) 
= \left[1+\frac{n(\bc)}{n(\bc^\prime)} e^{H(\bc)-H(\bc^\prime)}
\right]^{-1}
\label{eq:acceptance_rewrite}
\end{eqnarray}
These probabilities (\ref{eq:acceptance_rewrite}) ensure the edge-swapping process evolves into the stationary state on $\Omega=\{\bc\in\{0,1\}^{N^2}|~\bk^{\rm in}(\bc)=\bk^{\rm in},~\bk^{\rm out}(\bc)=\bk^{\rm out}\}$ defined by $p_\infty(\bc)= Z^{-1}\exp[-H(\bc)]$. The full degree-degree correlation structure of a directed graph $\bc$  is captured by the joint degree distribution of connected nodes
\begin{eqnarray}
W(\vec{k},\vec{k}^\prime|\bc)=\frac{1}{N\bra k\ket}\sum_{ij}c_{ij}~
\delta_{\vec{k},\vec{k}_i(\bc)}\delta_{\vec{k}^\prime,\vec{k}_j(\bc)}
\label{eq:W}
\end{eqnarray}
with $\vec{k}=(k^{\rm in},k^{\rm out})$. The maximum entropy distribution on $\Omega$, viz. all directed graphs with prescribed in- and out-degree sequences, 
that has the distribution (\ref{eq:W}) imposed as a soft constraint, i.e. $\sum_{\bc\in\Omega}p(\bc)W(\vec{k},\vec{k}^\prime|\bc)=W(\vec{k},\vec{k}^\prime)$ for all $(\vec{k},\vec{k^\prime})$, is 
\begin{eqnarray}
p(\bc)&=& Z^{-1} \prod_i \delta_{\vec{k}_i, \vec{k}_i(\bc)}
\label{eq:correlated}
\\
&& \times\prod_{ij}\left[ \frac{\bra k\ket}{N}Q(\vec{k}_i, \vec{k}_j)\delta_{c_{ij},1}+\Big(1\!-\!\frac{\bra k\ket}{N}Q(\vec{k}_i, \vec{k}_j)\Big)\delta_{c_{ij},0}  \right]
\nonumber
\end{eqnarray}
(see \cite{Roberts11}), 
in which $Q(\vec{k},\vec{k}^\prime)=W(\vec{k},\vec{k}^\prime)/p(\vec{k})p(\vec{k}^\prime)$ and $p(\vec{k})=p(k^{\rm in},k^{\rm out})$.
It is now trivial, following \cite{Coolen09}, to ensure that our MCMC process evolves to the measure (\ref{eq:correlated}) by choosing $H(\bc)=-\log p(\bc)$ in the probabilities (\ref{eq:acceptance_rewrite}). This gives
\begin{eqnarray}
 A(\bc|\bc^\prime) 
&&\nonumber
\\
&& \hspace*{-11mm}=\left[1+\frac{n(\bc)}{n(\bc^\prime)} 
\prod_{ij}
\frac{\frac{\bra k\ket}{N}Q(\vec{k}_i, \vec{k}_j)c^\prime_{ij}+\Big(1\!-\!\frac{\bra k\ket}{N}Q(\vec{k}_i, \vec{k}_j)\Big)(1\!-\!c^\prime_{ij})  }
{\frac{\bra k\ket}{N}Q(\vec{k}_i, \vec{k}_j)c_{ij}+\Big(1\!-\!\frac{\bra k\ket}{N}Q(\vec{k}_i, \vec{k}_j)\Big)(1\!-\!c_{ij})  }
\right]^{-1}
\nonumber
\\
&=& \left[1+\frac{n(\bc)}{n(\bc^\prime)} 
\prod_{ij}\left(\frac{
\frac{\bra k\ket}{N}Q(\vec{k}_i, \vec{k}_j)}{
1\!-\!\frac{\bra k\ket}{N}Q(\vec{k}_i, \vec{k}_j)}
\right)^{c^\prime_{ij}-c_{ij}}
\right]^{-1}
\label{eq:general_A_corr}
\end{eqnarray}
If the proposed move is a {\it square edge swap}, it is characterized by four distinct nodes
$(i,j,k,\ell)$, and takes us from a graph $\bc^\prime$ with $c^\prime_{ij}c_{k\ell}^\prime\bar{c}^\prime_{kj}\bar{c}^\prime_{i\ell}=1$ 
to a new graph $\bc$ with $\bar{c}_{ij}\bar{c}_{k\ell}c_{kj}c_{i\ell}=1$ (leaving all other $N^2\!-4$ bond variables unaffected). For such moves the acceptance probabilities (\ref{eq:general_A_corr}) become
\begin{eqnarray}
 A_{\square}(\bc|\bc^\prime) 
&=& \left[1+\frac{n(\bc)}{n(\bc^\prime)} 
\frac{\left(\frac{N}{\bra k\ket Q(\vec{k}_k, \vec{k}_j)}\!-\!1\right)
\left(\frac{N}{\bra k\ket Q(\vec{k}_i, \vec{k}_\ell)}\!-\!1\right)}
{\left(\frac{N}{\bra k\ket Q(\vec{k}_i, \vec{k}_j)}\!-\!1\right)
\left(\frac{N}{\bra k\ket Q(\vec{k}_k, \vec{k}_\ell)}\!-\!1\right)}
\right]^{-1}
\label{eq:A_square}
\end{eqnarray}
For large $N$ we may choose to approximate this by
\begin{eqnarray}
A_{\square}(\bc|\bc^\prime) 
\approx 
 \left[1+\frac{n(\bc)}{n(\bc^\prime)} 
 \frac{Q(\vec{k}_i, \vec{k}_j)Q(\vec{k}_k, \vec{k}_\ell)}
 {Q(\vec{k}_k, \vec{k}_j)Q(\vec{k}_i, \vec{k}_\ell)}
\right]^{-1}
\end{eqnarray}
If the proposed move is a {\it triangle edge swap}, it is characterized by three distinct nodes
$(i,j,k)$, and takes us from a graph $\bc^\prime$ with $c^\prime_{ij}c_{jk}^\prime c^\prime_{ki}\bar{c}^\prime_{ji}\bar{c}_{kj}^\prime \bar{c}^\prime_{ik}=1$ 
to a new graph $\bc$ with $\bar{c}_{ij}\bar{c}_{jk} \bar{c}_{ki}c_{ji}c_{kj} c_{ik}=1$ 
 (leaving all other $N^2\!-6$ bond variables unaffected). Now the acceptance probabilities (\ref{eq:general_A_corr}) become
 \begin{eqnarray}
 A_{\triangle}(\bc|\bc^\prime) 
&=& \label{eq:A_triangle}
\\
&&\hspace*{-13mm}
 \left[1+\frac{n(\bc)}{n(\bc^\prime)} 
\frac{\left(\frac{N}{\bra k\ket Q(\vec{k}_j, \vec{k}_i)}\!-\!1\right)
\left(\frac{N}{\bra k\ket Q(\vec{k}_k, \vec{k}_j)}\!-\!1\right)
\left(\frac{N}{\bra k\ket Q(\vec{k}_i, \vec{k}_k)}\!-\!1\right)}
{\left(\frac{N}{\bra k\ket Q(\vec{k}_i, \vec{k}_j)}\!-\!1\right)
\left(\frac{N}{\bra k\ket Q(\vec{k}_j, \vec{k}_k)}\!-\!1\right)
\left(\frac{N}{\bra k\ket Q(\vec{k}_k, \vec{k}_i)}\!-\!1\right)}
\right]^{-1}\hspace*{-5mm}
\nonumber
\end{eqnarray}
For large $N$ we may choose to approximate this by
 \begin{eqnarray}
 A_{\triangle}(\bc|\bc^\prime) 
&=& \left[1+\frac{n(\bc)}{n(\bc^\prime)} 
\frac{Q(\vec{k}_i, \vec{k}_j)Q(\vec{k}_j, \vec{k}_k)Q(\vec{k}_k, \vec{k}_i)}
{Q(\vec{k}_j, \vec{k}_i)Q(\vec{k}_k, \vec{k}_j) Q(\vec{k}_i, \vec{k}_k)}
\right]^{-1}
\end{eqnarray}

\section{Numerical simulations of the canonical randomization process}

\noindent
In this section we describe numerical simulations of our canonical MCMC graph randomization process and its `accept all' edge swapping counterpart, applied to synthetic networks and to biological signalling networks.  
The most convenient marker of sampling bias in randomisation is the mobility $n(\bc)$ itself, which we will therefore use as to monitor the dynamics of the process. We used the Mersenne Twister random number generator from \cite{MersenneTwister}.  
For numerical implementation, we use expressions for the incremental change in the mobility terms following  a single edge swap move (similar to how this was done for nondirected networks \cite{Coolen09}) -- see appendix \ref{app:mobility_change}. This avoids having to calculate $n(\bc)$ after each move, which would involve repeated matrix multiplications. Full source code (in C++) and Windows executables of our implementation are available on request.

\subsection{Split flow network}

\begin{figure}[h]
  \begin{center}
    \includegraphics[width=0.8\linewidth]{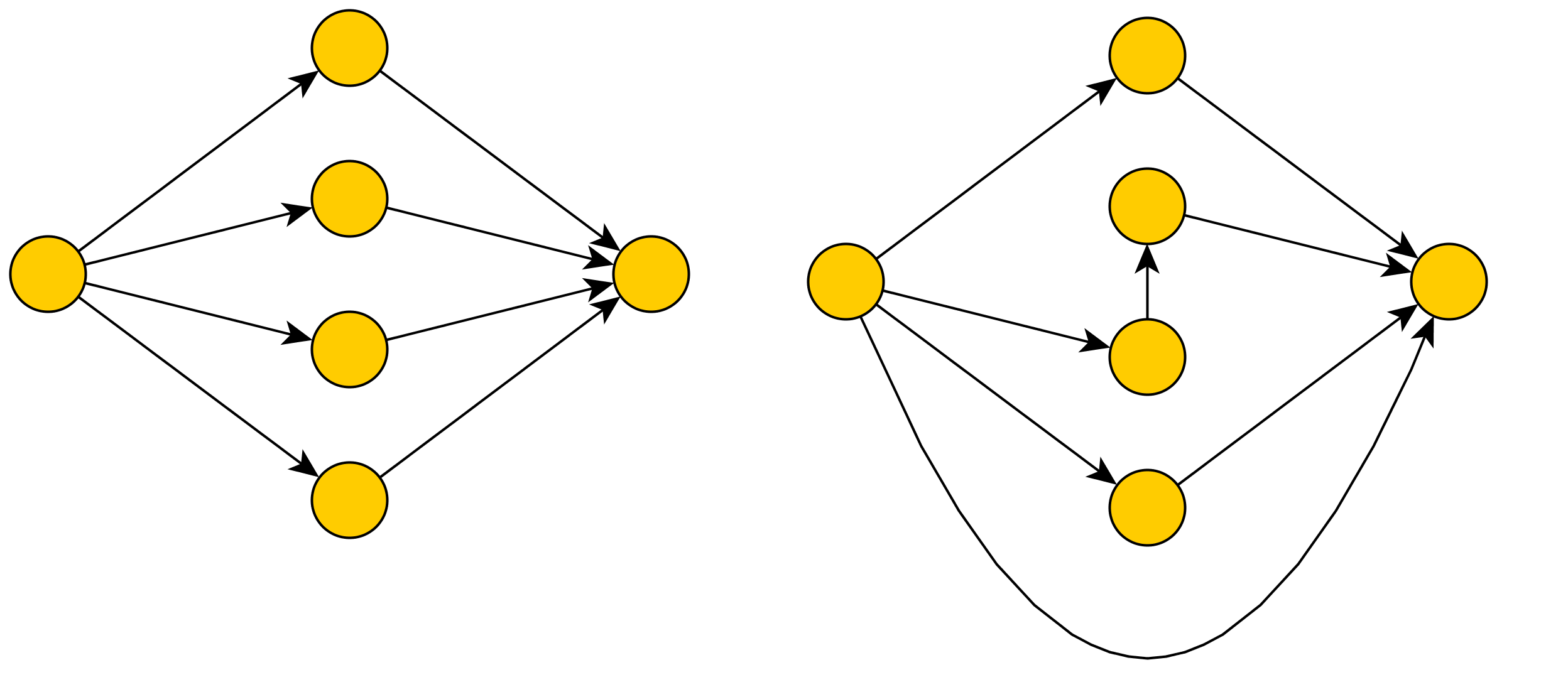}
  \end{center}
    \vspace*{-3mm}
  \caption{The possible realisations of a split flow type network, with $N=K+2$. The left hand configuration has a mobility of $K(K-1)$; there is only one such configuration. The right hand configuration has mobility of $2K-3$; there are $K(K-1)$ such configurations. }
  \label{fig:split-flow}
\end{figure}
\begin{figure}[t]
\vspace*{-5mm}
\hspace*{-6mm}
\unitlength=0.21mm
\begin{picture}(350,350)
\put(0,40){\includegraphics[width=420\unitlength]{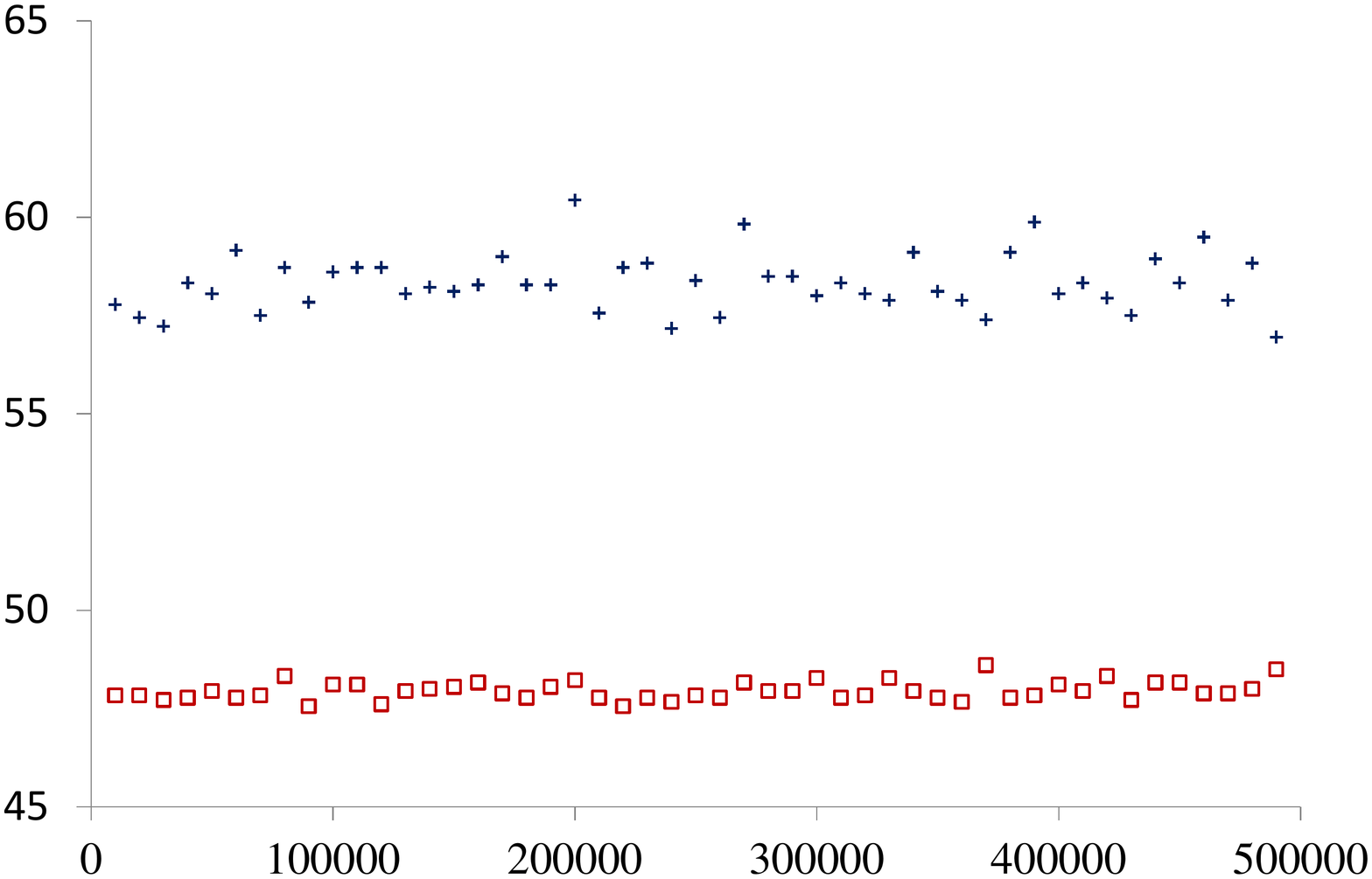}}
\put(190,45){\it iterations} \put(-20,220){$\bra n(\bc) \ket$}
\end{picture}
   \vspace*{-7mm}
   \caption{Comparison for `split flow' networks with $K=25$ of randomization via `accept all' edge swapping (squares) versus edge swapping with the canonical acceptance probabilities (crosses). The mobility $\bra n(\bc) \ket$ is used as a dynamical observable, since its expectation value is sensitive to sampling bias. Each marker gives the average mobility over 10,000 iterations. Observed values are in good agreement with theoretical predictions: $ \bra n(\bc) \ket\approx 58.32$ for `accept all' edge swapping (predicted:  58.52),  versus $\bra n(\bc) \ket\approx  47.95$ for correct edge swapping (predicted:  47.92).  }
\label{fig:example2_trajectory}
\end{figure}

\noindent
A split flow network, see e.g.  \cite{Milo04}, is built as follows. Node $i=1$ has degrees $(k^{\rm in}_1,k^{\rm out}_1)=(0,K)$,
we have $K$ nodes ($i=2\ldots K+1$) with degrees $(k^{\rm in}_i,k^{\rm out}_i)=(1,1)$, and a final node with degrees $(k^{\rm in}_{K+2},k^{\rm out}_{K+2})=(K,0)$. There exist two types of graph with this specified degree sequence. The first is shown in the left of figure   \ref{fig:split-flow}. The second type is obtained from the first by choosing two of the $K$ `inner nodes', of which one will cease to receive a link from $i=1$ and the second will cease to provide a link to $i=K+2$; so the mobility of the left graph is $n(\bc)=K(K-1)$. On the right-hand side configurations in figure   \ref{fig:split-flow} we can execute three possible square edge swap types: returning to the previous state (1 move), changing the internal node that is not receiving a link from $i=1$ ($K-2$ moves), or changing the internal node that is not sending a link to $i=K+2$ ($K-2$ moves), giving a total mobility for the graphs on the right of $n(\bc)=2K-3$. The total number of such split flow networks is $|\Omega|=K(K-1)+1$. 

Figure \ref{fig:example2_trajectory} shows graph randomisation dynamics for a split-flow network with $K=25$, comparing `accept all' edge swapping (which would sample graphs with the bias  $p(\bc)=n(\bc)/\sum_{\bc^\prime\in\Omega}n(\bc^\prime)$) 
to the canonical edge swap process (\ref{eq:togetflat}) that is predicted to give unbiased sampling of graphs $p(\bc)=1/|\Omega|$. 
The predicted expectation values of the mobilities in the two sampling protocols are 
\begin{eqnarray*}
{\rm `accept~all'\!:}&~~~\bra n(\bc)\ket=\frac{\sum_{\bc\in\Omega}n^2(\bc)}{\sum_{\bc\in\Omega}n(\bc)}&
= \frac{5K^2\!-\!13K\!+\!9}{2(K\!-\!1)}\approx 58.52
\\
{\rm canonical\!:}&~~~\bra n(\bc)\ket=\frac{\sum_{\bc\in\Omega}n(\bc)}{|\Omega|}&
=\frac{2K(K\!-\!1)^2}{1\!+\!K(K\!-\!1)}~~~\approx 47.92
\end{eqnarray*}
 The simulation results confirm these quantitative predictions (see caption of figure \ref{fig:example2_trajectory} for details), and 
 underline the sampling bias caused by `accept all' edge swapping, as well as the lack of such a bias in our canonical MCMC process.

\subsection{Nearly hardcore networks}
\label{section:nearly_hardcore}

\begin{figure}[t]
\setlength{\intextsep}{0pt} 
  \begin{center}
    \includegraphics[width=0.15\textwidth]{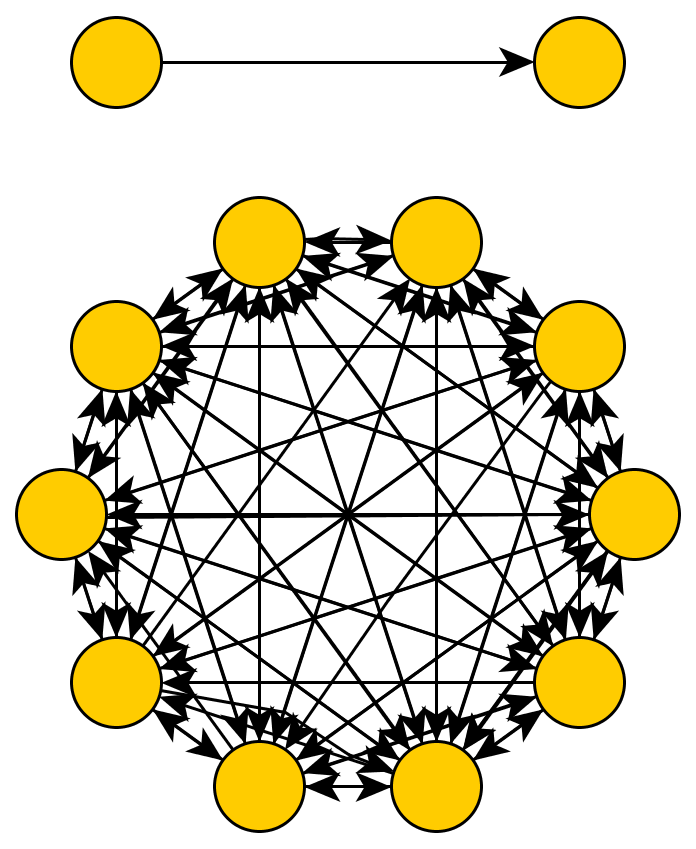}
  \end{center}
  \vspace*{-4mm}
  \caption{The directed version of a `nearly hardcore' network. Given the imposed degree sequences, there are only two types of graphs: 
  the one shown here, and the one obtained by via an edge swap that involves the nodes of the isolated link and two nodes from the core. }
  \label{fig:nhc}
\end{figure}

\begin{figure}[t]
\vsp
  \begin{center}
    \includegraphics[width=0.47\textwidth]{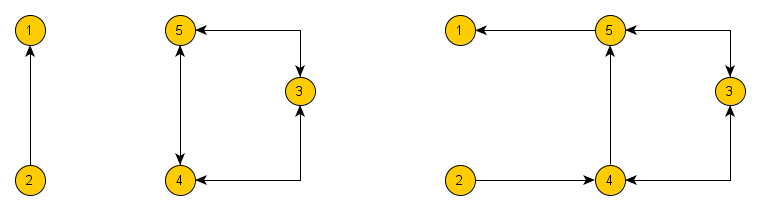}
  \end{center}
  \vspace*{-3mm}
  \caption{Illustration of the edge swap that transforms a nearly hardcore graph from state A to one of the type B states. }
  \label{fig:AtoB}
\end{figure}

\noindent
`Nearly hardcore' networks are another example of graphs for which `accept all' edge swap sampling are known to exhibit a significant bias  \cite{Coolen09}. The directed version of such networks is constructed from a single isolated bond plus  a complete subgraph of size $K=N-2$. 
See figure \ref{fig:nhc}. Triangle swaps are not possible. 
 From the graph shown in the figure (the `mobile' state, A) there are $K(K\!-\!1)$ ways to choose two nodes of the core to combine with the two non-core nodes to form an edge swap quartet, hence this state has $n_A(\bc)=K(K\!-\!1)$. 
After an edge swap the graph in  figure \ref{fig:nhc} is replaced by one in which one non-core node receives a link from the core, and the other sends a link to the core; see figure \ref{fig:AtoB}. There are $K(K\!-\!1)$ such graphs, to be called type B, hence the total number of nearly hardcore graphs is $|\Omega|=K(K\!-\!1)\!+\!1$. From each type B graph the inverse swap can be applied, plus $2(K\!-\!2)$ further moves that each equate to replacement of one of the core nodes involved in the previous swap by another. Hence $n_B(\bc)=2K\!-\!3$. These statements are confirmed by formula (\ref{eq:square_mobility_term}). 

\begin{figure}[t]
\vspace*{-5mm}
\hspace*{-6mm}
\unitlength=0.21mm
\begin{picture}(350,350)
\put(0,40){\includegraphics[width=420\unitlength]{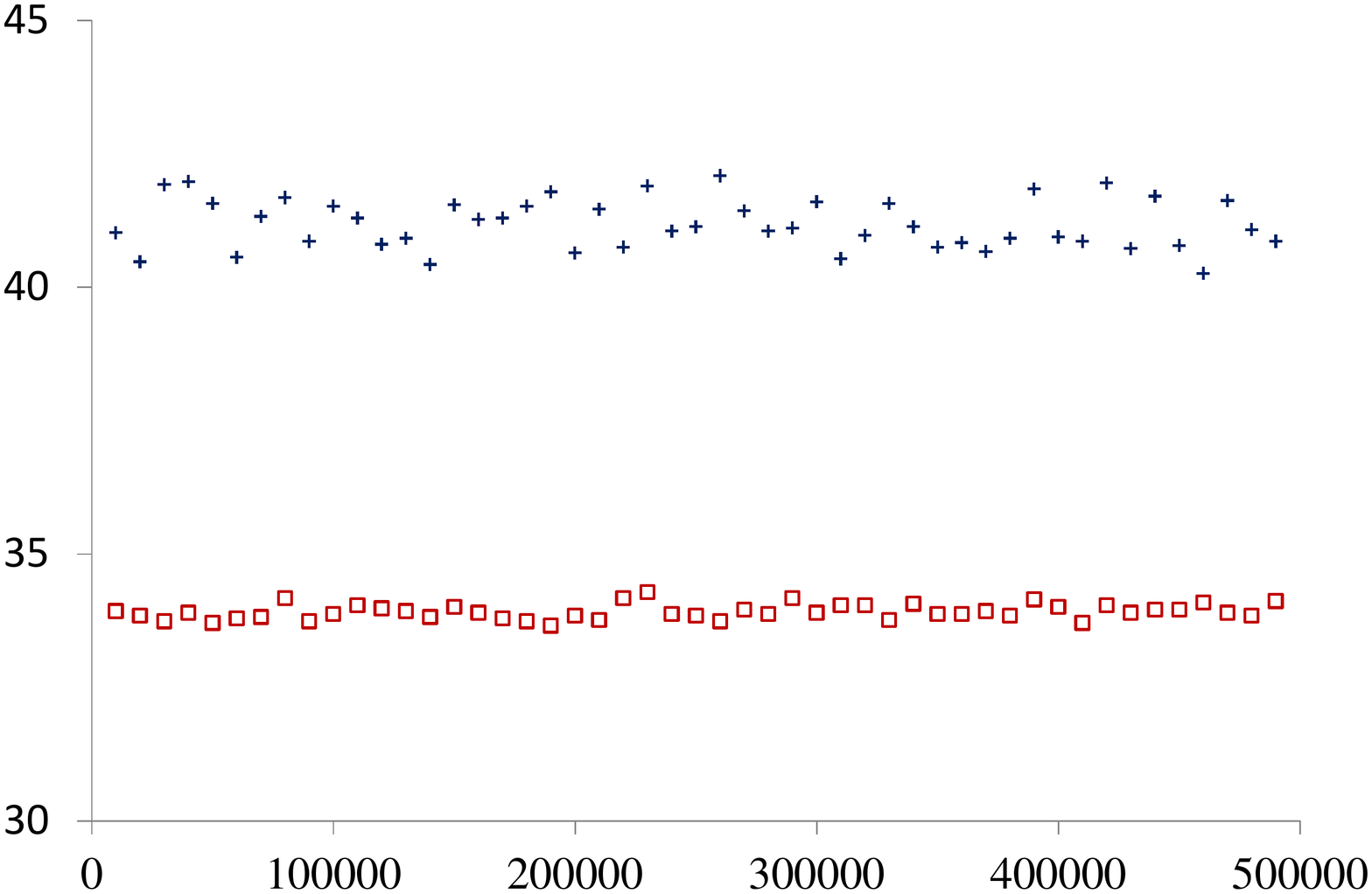}}
\put(190,45){\it iterations} \put(-17,200){$\bra n(\bc) \ket$}
\end{picture}
   \vspace*{-9mm}
   \caption{
   Comparison for `nearly hardcore' networks with $K=18$ of randomization via `accept all' edge swapping (squares) versus edge swapping with the canonical acceptance probabilities (crosses). 
   Each marker gives the average mobility over 10,000 iterations. 
 Observed mobility values are again in good agreement with theoretical predictions: $ \bra n(\bc) \ket \approx 41.09$ for `accept all'  (predicted: 41.03),  versus $\bra n(\bc) \ket \approx   33.92 $ for correct edge swapping (predicted: 33.89).  }
 \label{fig:hardcore20}
\end{figure}

The predicted expectation values of the mobilities in the two sampling protocols,
`accept all' edge swapping (which would sample graphs with the bias  $p(\bc)=n(\bc)/\sum_{\bc^\prime\in\Omega}n(\bc^\prime)$) and 
the canonical edge swap process (\ref{eq:togetflat}) (predicted to give unbiased sampling of graphs $p(\bc)=1/|\Omega|$), 
 are 
\begin{eqnarray*}
{\rm `accept~all'\!:}&~~~\bra n(\bc)\ket&
= \frac{n_A^2(\bc)\!+\!K(K\!-\!1)n_B^2(\bc)}{n_A(\bc)\!+\!K(K\!-\!1)n_B(\bc)}
= 
\frac{5K^2\!-\!13K\!+\!9}{2(K\!-\!1)}
\\
{\rm canonical\!:}&~~~\bra n(\bc)\ket&
=  \frac{n_A(\bc)+K(K\!-\!1)n_B(\bc)}{1\!+\!K(K\!-\!1)}
=  \frac{2K(K\!-\!1)^2}{1\!+\!K(K\!-\!1)}
\end{eqnarray*}
Figure  \ref{fig:hardcore20}  shows graph randomisation dynamics for a nearly hardcore network with $K=18$ (so $N=20$). Here the theory, i.e. the previous two formulae, predicts that we should  see $\bra n(\bc)\ket\approx 41.03$ for `accept all' edge swapping, and 
$\bra n(\bc)\ket\approx 33.89$ for unbiased sampling.
  Again  the simulation results confirm our predictions (see caption of figure \ref{fig:hardcore20} for details).

\subsection{Application to gene regulation networks}

\begin{figure}[t]
\vspace*{-5mm}
\hspace*{-6mm}
\unitlength=0.21mm
\begin{picture}(350,350)
\put(-10,20){\includegraphics[width=460\unitlength]{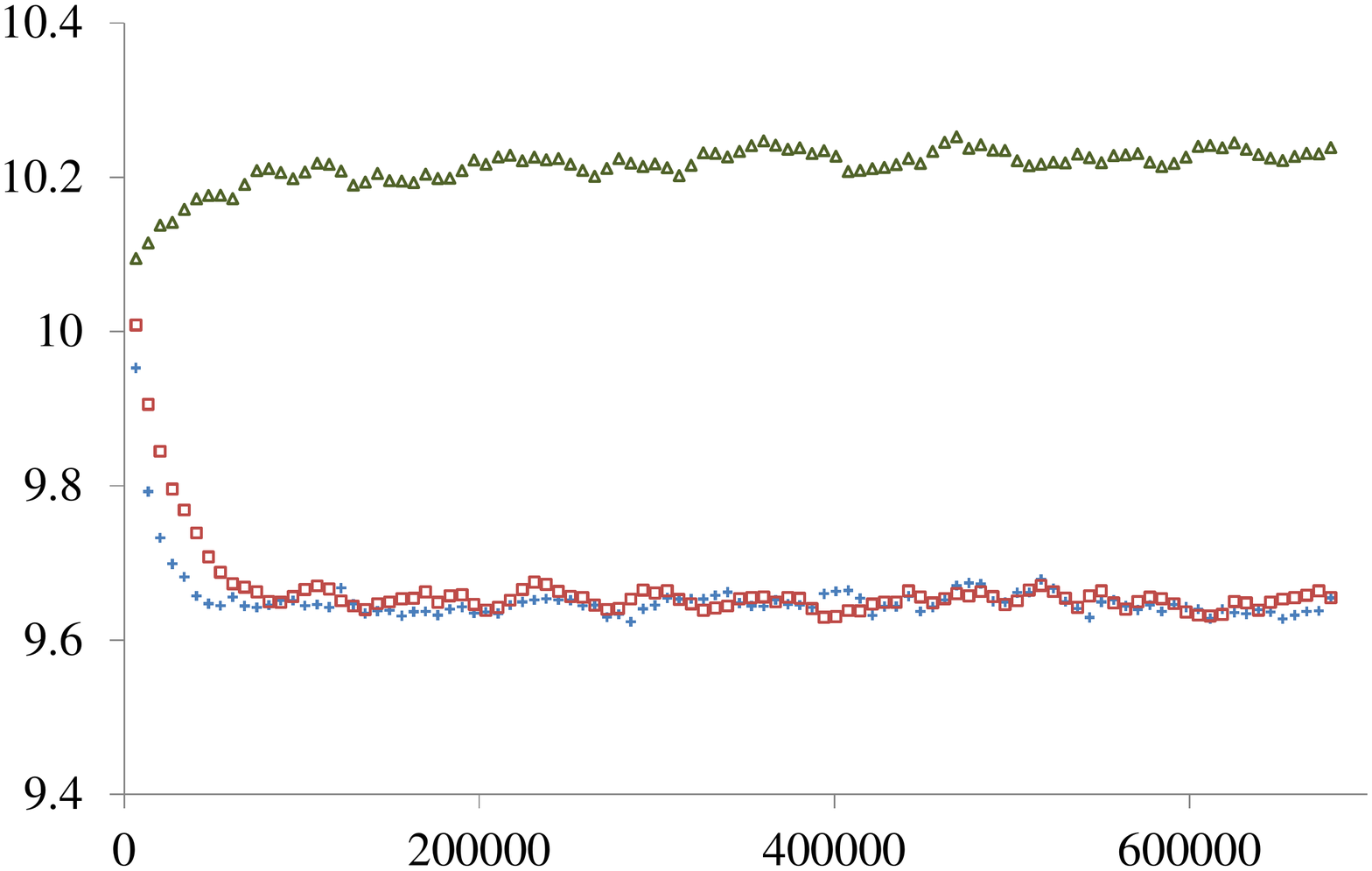}}
\put(190,45){\it iterations} \put(-15,200){$\frac{\bra n(\bc)\ket}{N^2}$}
\end{picture}
   \vspace*{-9mm}
\caption{Randomization dynamics for the gene regulation network data of \cite{Hughes}. The observable shown is a rolling average of the normalized average square mobility 
$\bra n_{\square}(\bc)\ket/N^2$.  
We compare 
`accept all' edge swapping ($+$), canonical edge swapping aimed at uniform sampling of all graphs with the biological degree sequence  of the biological network ($\square$), and canonical 
 edge swapping aimed at uniform sampling of all graphs with the degree sequence $(\vec{k}_1,\ldots,\vec{k}_N)$ {\em and} the degree-degree correlation kernel $W(\vec{k},\vec{k}^\prime)$ of the biological network ($\triangle$). Hamming distances between the start and end networks of $\square$, $+$ and $\triangle$ were 0.8, 0.8 and 0.75 respectively.
}
\label{fig:hughes}
\end{figure}

\begin{figure}[h]
\vspace*{-5mm}
\hspace*{-6mm}
\unitlength=0.21mm
\begin{picture}(350,355)
\put(-16,25){\includegraphics[width=450\unitlength]{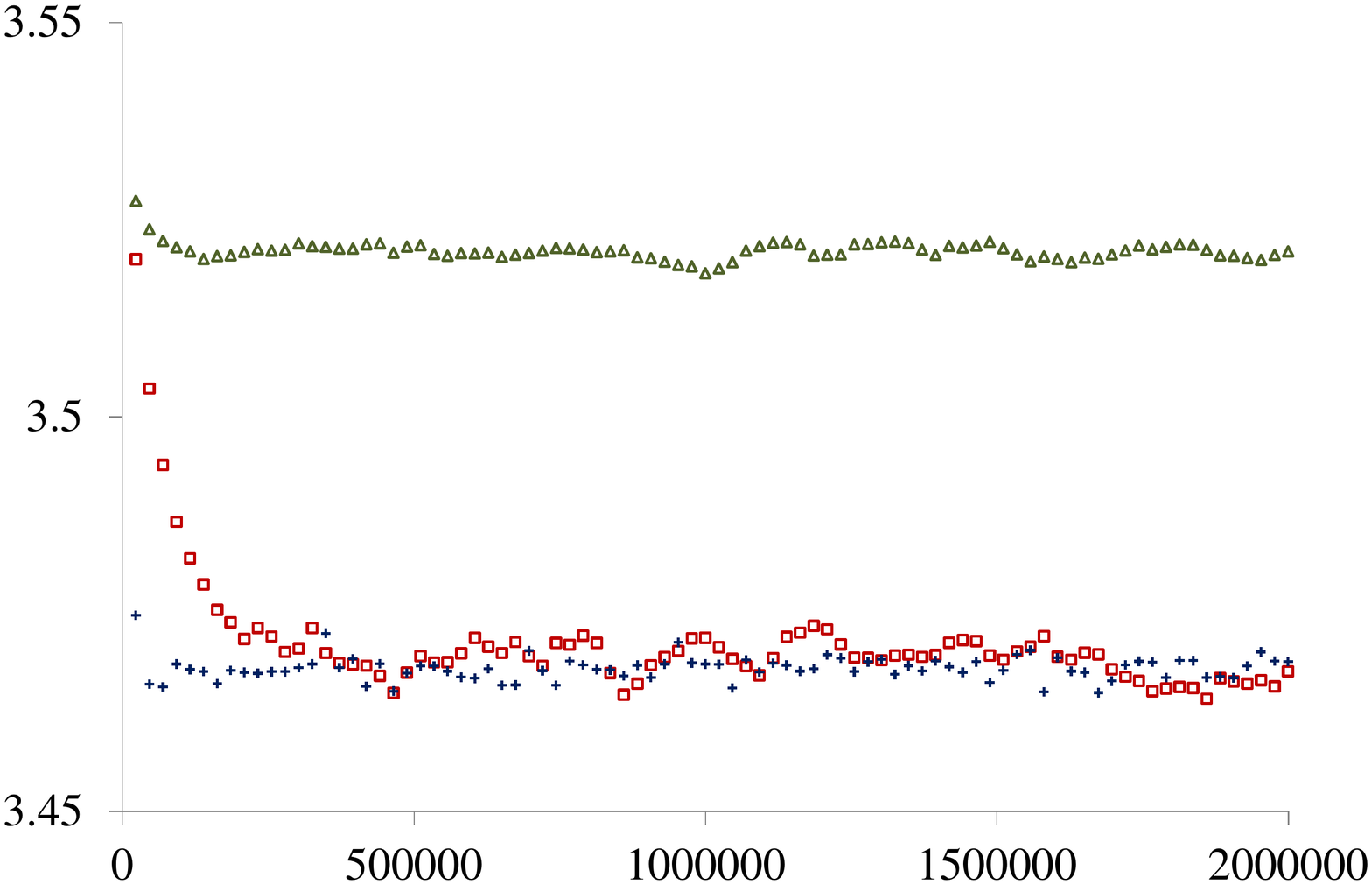}}
\put(190,40){\it iterations} \put(-15,200){$\frac{\bra n(\bc)\ket}{N^2}$}
\end{picture}
   \vspace*{-7mm}
\caption{Randomization dynamics for the gene regulation network of \cite{Harbison}. The key and the axes are the same as in figure \ref{fig:hughes}. Hamming distances between the start and end networks of $\square$, $+$ and $\triangle$ were 0.94, 0.94 and 0.86 respectively.
}
\label{h202LO}
\end{figure}

\noindent 
Gene regulation networks are important examples of directed biological networks. 
Figures \ref{fig:hughes} and \ref{h202LO} show numerical results of the randomization dynamics 
 applied to the gene regulation network data of \cite{Hughes} (with  $N=5654$ nodes) and \cite{Harbison} (with  $N=3865$ nodes), respectively. We apply all three randomization processes discussed so far in this paper, viz. `accept all' edge swapping, canonical edge swapping aimed at uniform sampling of all graphs with the degree sequences of the biological network, and canonical 
 edge swapping aimed at uniform sampling of all graphs with the degree sequence $(\vec{k}_1,\ldots,\vec{k}_N)$ {\em and} (on average) the degree-degree correlation kernel $W(\vec{k},\vec{k}^\prime)$ of the biological network. 
In contrast to the synthetic examples in the previous subsection, 
in gene regulation networks we do not observe significant divergence between `accept all' versus canonical edge swap randomization; this is similar to what was observed earlier for the randomization of protein-protein interaction networks in \cite{Coolen09}. 
We also see that in both cases the biological network is significantly more mobile than the typical network with the same degree sequence. 
However,  figure \ref{fig:hughes} suggests that the set of networks that share with the biological one both the degree sequence {\em and} the degree correlations (and hence resemble more closely the biological network under study) all have high mobilities.

Implementating degree-degree correlation targeting directly has the effect of severely reducing the space of graphs through which the process can pass, hence we would expect finite-size effects to be more pronounced. The process would be less restricted, and hence more natural, with a smoothed target degree-degree correlation. There is a trade-off between the flexibility of the process and the accuracy of the targeting. 
We have used a light Gaussian smoothing, generalising what was used in \cite{Fernandes10} to the higher dimension we need. The best choice target degree-degree correlations - including decisions about smoothing - will very much depend on the particular problem being studied. 

\subsection{Targeting degree-degree correlation}

\begin{figure}[h]
\vspace*{0mm}
\hspace*{0mm}
\unitlength=0.23mm
\begin{picture}(300,600)
\put(-40,0){\includegraphics[width=200\unitlength]{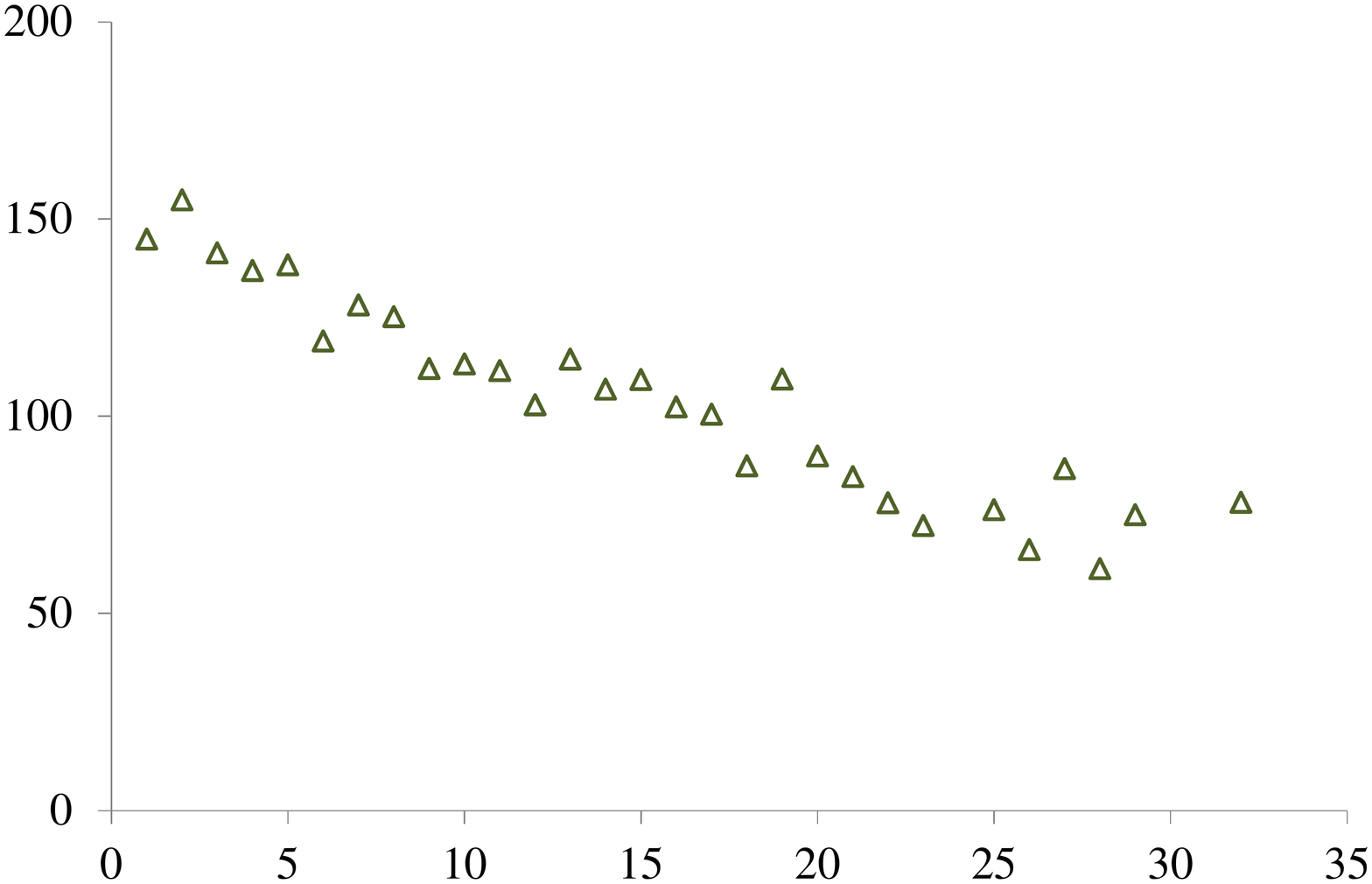}}
\put(150,0){\includegraphics[width=200\unitlength]{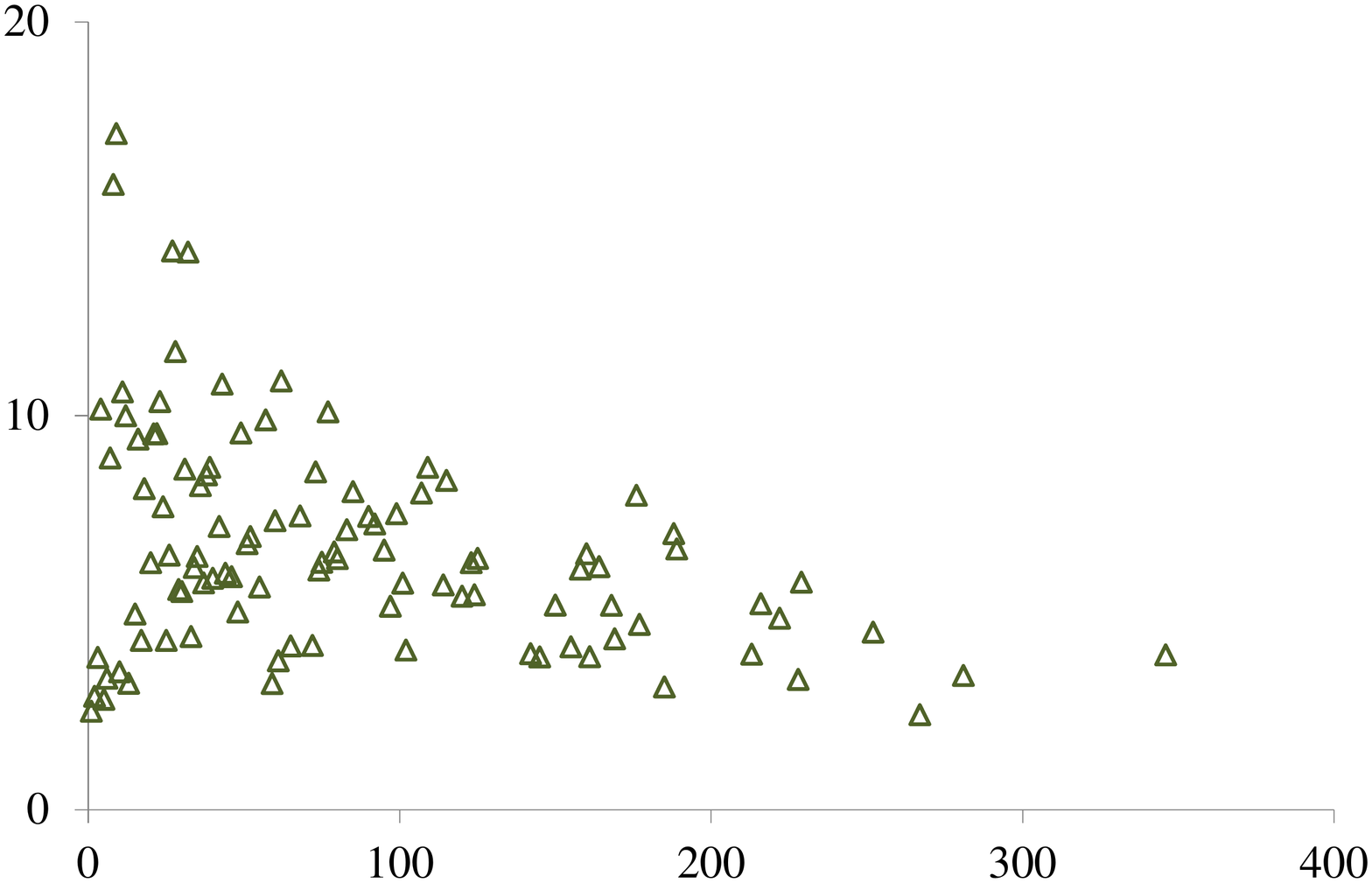}}
\put(60,1){\footnotesize $k^{\rm in}$}  \put(250,1){\footnotesize $k^{\rm out}$}
\put(120,174){\small\it Final network}
\put(65,161){\small\it (target: preserved degree correlations)}
 \put(-30,135){\footnotesize $\bra k^{\rm out}_{\rm nn} \ket_{\rm in}$}
 \put(160,135){\footnotesize $\bra k^{\rm in}_{\rm nn} \ket_{\rm out}$}

\put(-40,210){\includegraphics[width=200\unitlength]{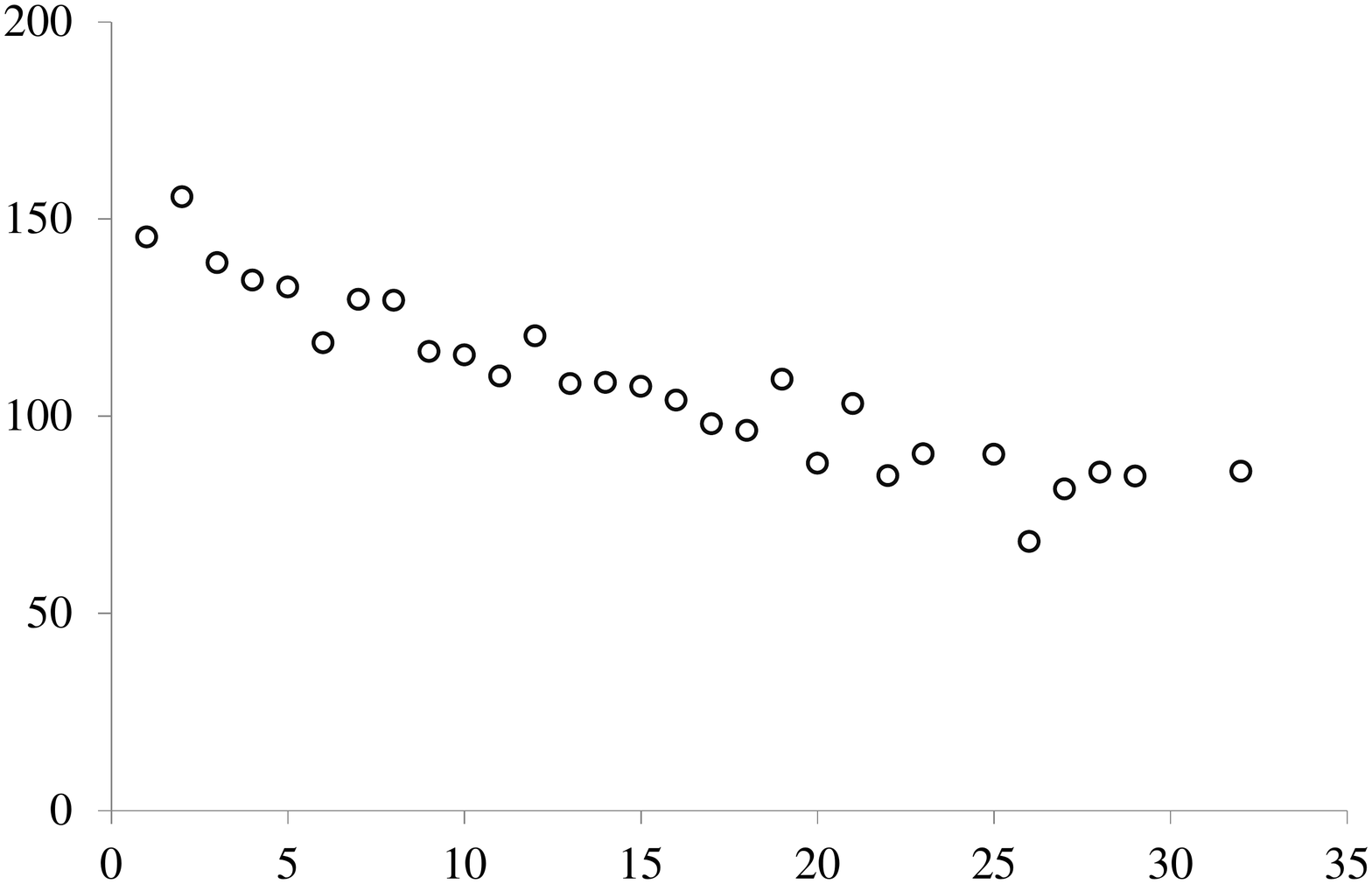}}
\put(150,210){\includegraphics[width=200\unitlength]{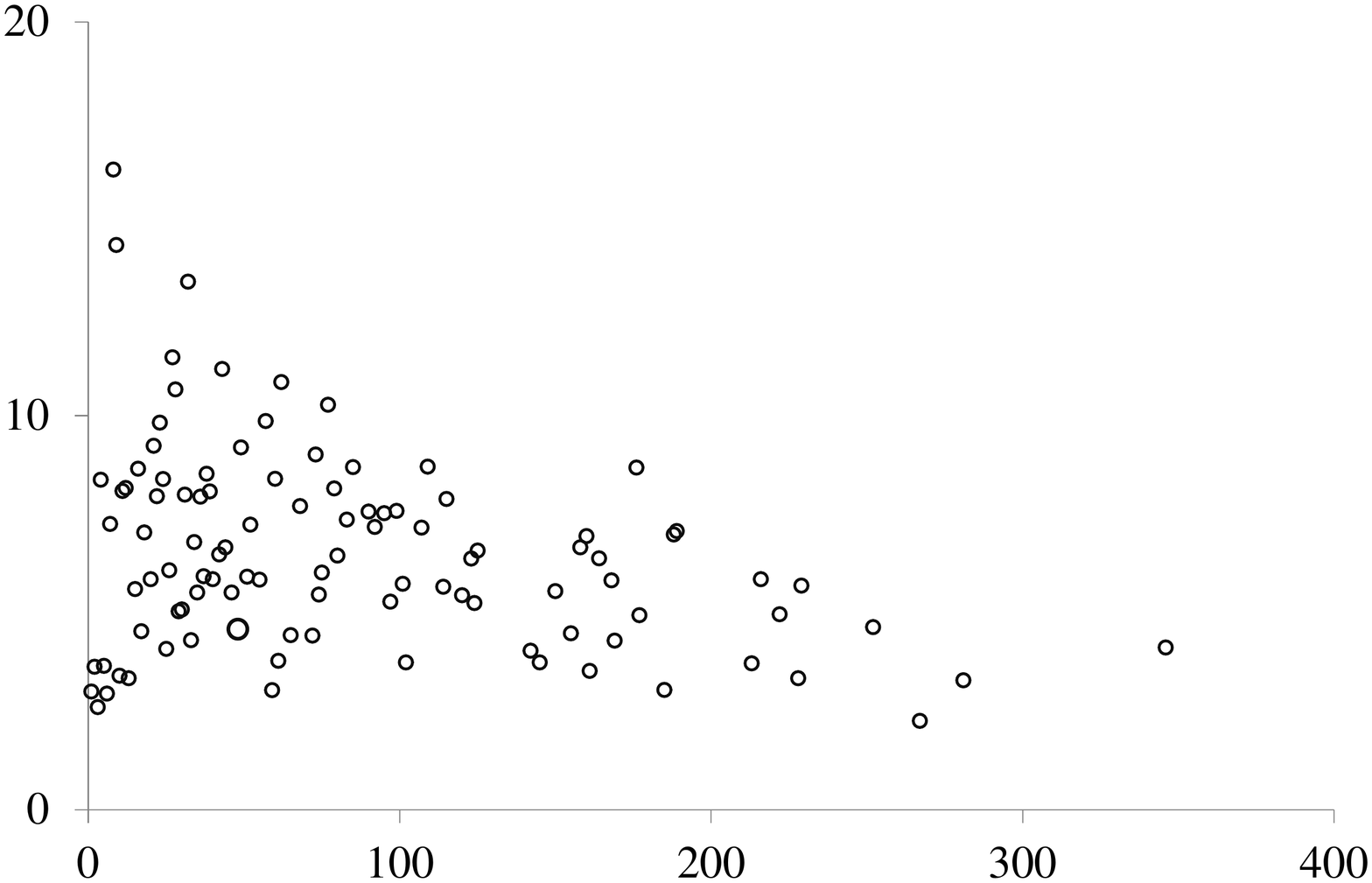}}
\put(60,211){\footnotesize $k^{\rm in}$}  \put(250,211){\footnotesize $k^{\rm out}$}
\put(105,375){\small\it Biological network }
 \put(-30,345){\footnotesize $\bra k^{\rm out}_{\rm nn} \ket_{\rm in}$}
  \put(160,345){\footnotesize $\bra k^{\rm in}_{\rm nn} \ket_{\rm out}$}

\put(-40,414){\includegraphics[width=200\unitlength]{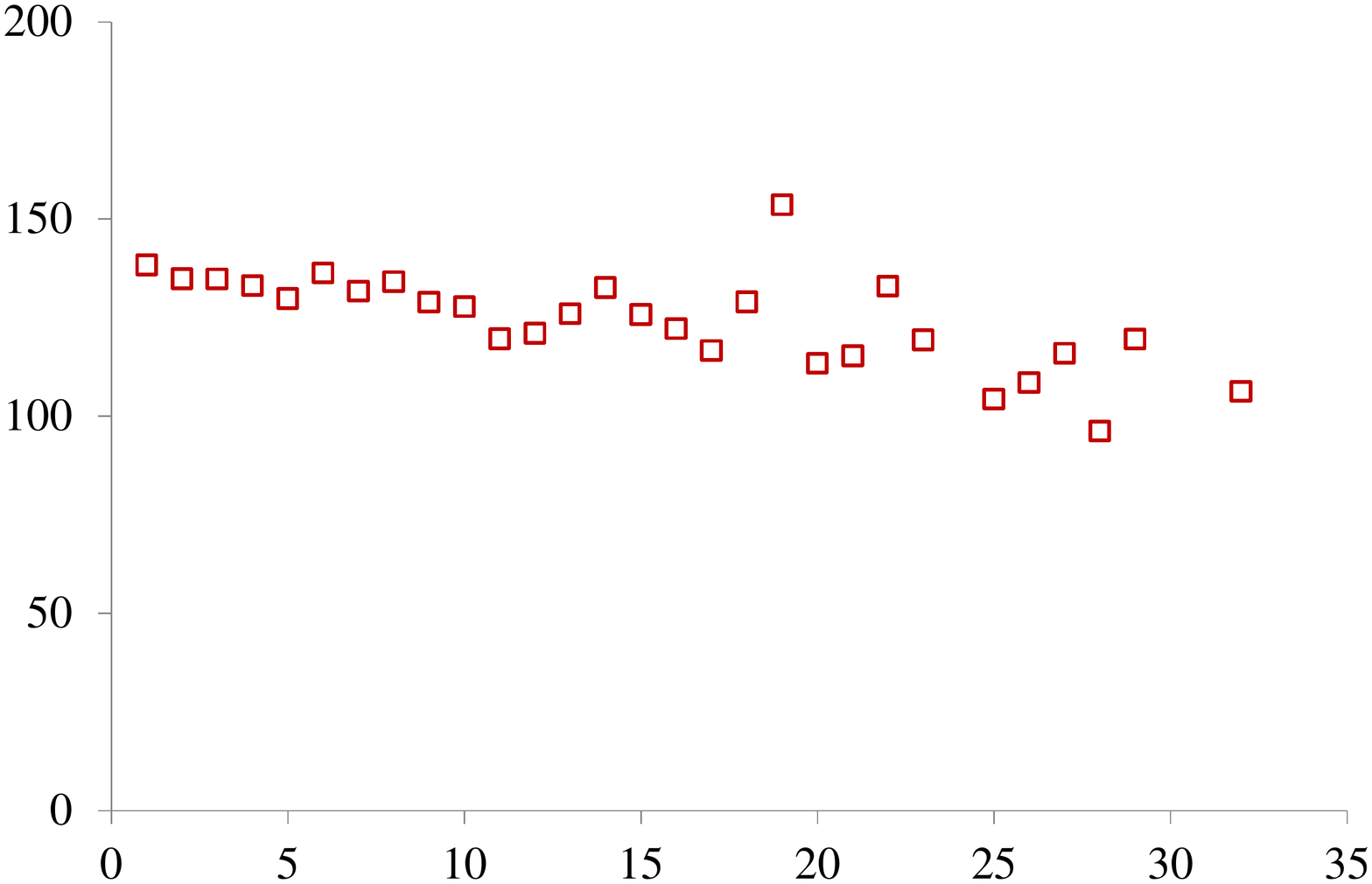}}
\put(150,410){\includegraphics[width=200\unitlength]{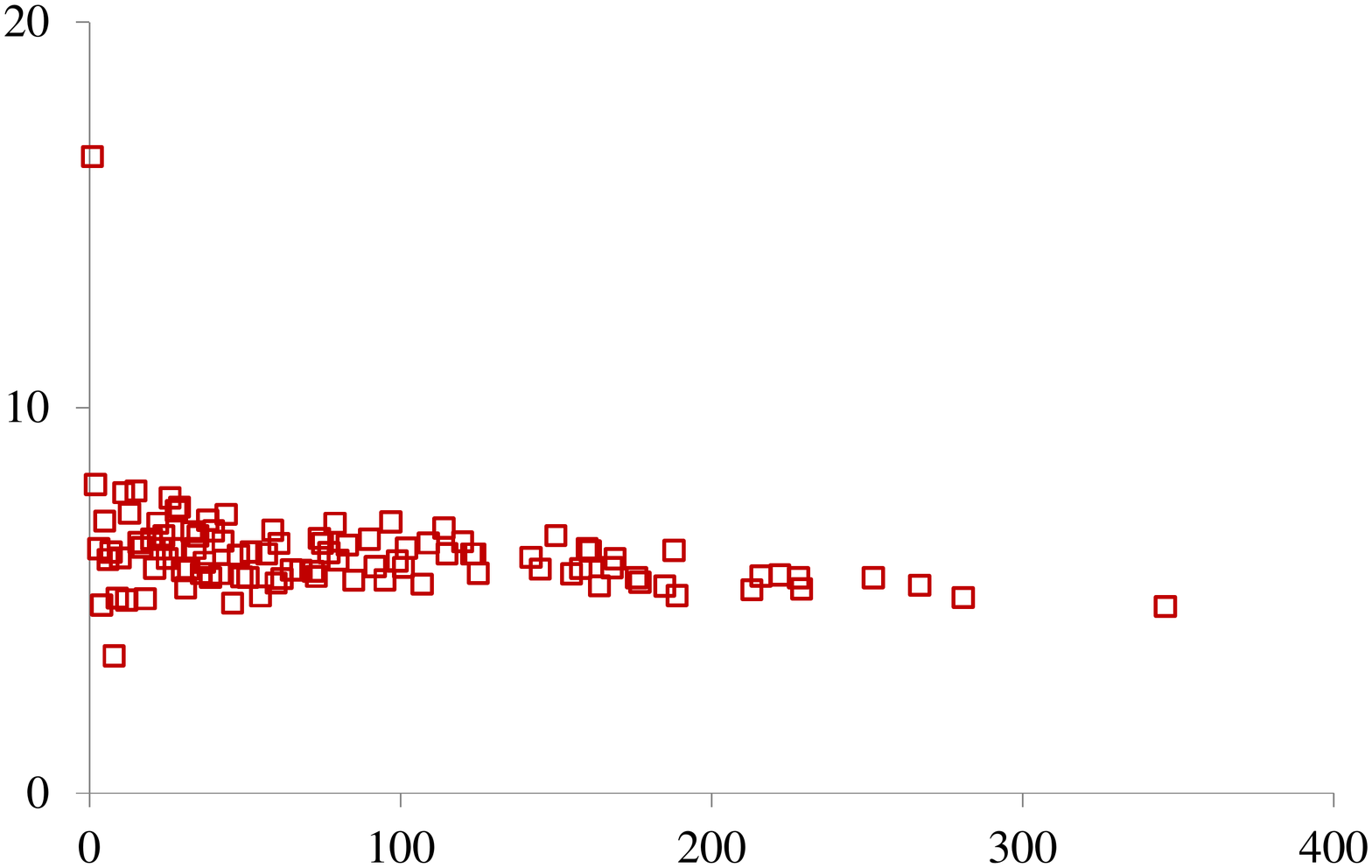}}
\put(60,411){\footnotesize $k^{\rm in}$}  \put(250,411){\footnotesize $k^{\rm out}$}
\put(75,582){\small\it Final network (target: flat measure)}
 \put(-30,549){\footnotesize $\bra k^{\rm out}_{\rm nn} \ket_{\rm in}$}
    \put(160,549){\footnotesize $\bra k^{\rm in}_{\rm nn} \ket_{\rm out}$}
 
\end{picture}

\caption
{
These charts summarize the degree-degree correlations observed in the original
network (middle row $\circ$),
 the final network after the process targeting the flat measure (top row $\square$) 
and the process tailored to preserve the degree-degree correlation pattern of the original network (bottom row $\triangle$). The data used is based on \cite{Harbison} and the process shown in figure \ref{h202LO}.
The left hand charts summarize the correlation  between the in-degree of a node and the average out-degree  $\bra k^{\rm out}_{\rm nn} \ket_{\rm in}$ of its in-neighbours. The right hand charts summarize the correlation between the out-degree of a node and the average in-degree  $\bra k^{\rm in}_{\rm nn} \ket_{\rm out}$ of its out-neighbours. 
This representation was chosen as a widely adopted and easy to interpret measure of the assortativity of a directed network. 
}
\label{fig:correlation}
\end{figure}

\noindent
In addition to being unbiased, the canonical MCMC process in this paper can sample according to any specified measure on the space of degree-constrained graphs. The particular example which we've developed is the generation of directed graphs from the tailored ensemble (\ref{eq:correlated}), via the acceptance probabilities (\ref{eq:A_square},\ref{eq:A_triangle}). Figures \ref{fig:hughes} and \ref{h202LO} show the trajectory of this process for two different datasets. Figure \ref{fig:correlation} is provided to illustrate that the network corresponding to this process successfully reproduces the key features of the assortativity of the real network. In particular, the characteristic downwards slope was postulated by \cite{Maslov02} to be a key feature of protein networks, associated with greater stability and improved specificity.  

%
%
%
%
%
%
%
%
%
%

\section{Conclusion}

\noindent
In this paper we have built on the work of \cite{Rao96} and \cite{Coolen09} to define an ergodic and unbiased stochastic process for randomising directed binary non-self-interacting networks,  which keeps the number of in- and out- connections of each node constant. 
The result takes the form of a canonical Markov Chain Monte Carlo (MCMC) algorithm based on simple direct edge swaps and triangle reversals, with nontrivial move acceptance probabilities that are calculated from the current state of the network only. The acceptance probabilities correct for the entropic bias in `accept all' edge-swap randomization, which is caused by the state dependence of the number of moves that can be executed (the `mobility' of a graph). 

Our process is precise for any network size and network topology, and sufficiently versatile to allow random directed graphs 
with the correct in- and out-degree sequence to be generated with arbitrary desired sampling probabilities. The algorithm can be used e.g. to generate truly unbiased random directed graphs with imposed degrees for hypothesis testing (in contrast to the `edge stub' algorithm or the `accept all' edge swap algorithm, both of which are biased), or to generate more sophisticated null models which inherit from a real network both the degree sequence {\em and} the degree correlations, but are otherwise random and unbiased. 

Our core insight is similar to \cite{Artzy-Randrup} and \cite{Rao96}. However, our work takes the formalism further, and generates a direct adjustment to the MCMC based on the current state of the network only, rather than a retrospective adjustment to the observed process \cite{Artzy-Randrup} or a search of the entire state-space \cite{Rao96}. Moreover, our approach can be generalised to generate more tailored null-models (e.g. our example of targeting a specified degree-degree correlation).

 We have derived bounds to predict for which degree sequences the differences between `accept all' and correct randomization (i.e. the effects of sampling bias)  are negligible.  
Application to  synthetic networks showed a large discrepancy between the `accept all' and correct randomization processes, 
and good agreement with our theoretical predictions for the values of key observables that are affected by the entropic bias of incorrect randomization. 
 For the biological networks which we studied (gene regulation networks) we find the differences between correct and incorrect sampling in the space of graphs with imposed degree sequences to be 
 negligible.  However, this cannot be relied upon to continue in future studies, especially when network datasets become less sparse, or randomization processes which target more complicated topological observables are used. 
 
 Biological signalling networks tend to have `fat-tailed' distributions with low average degree and relatively high clustering levels, whereas in a graph ensemble defined by prescribing  in- and out- degree sequences and  uniform graph probabilities, graphs will typically have $\order(1)$ triangles per node or less. Hence, if we run edge swap processes on such ensembles, by the time equilibration is approached the algorithm will typically be moving through networks with low clustering, where the change in mobility coming from those terms that `count' triangles will be very low. However, this will be different if we target a non-flat measure, for instance if we generate graphs with degree-degree correlations. Since biological degree-degree correlations seem to be associated with clustering,  it will become increasingly dangerous to assume that  the sampling bias caused by using `accept all' edge swap dynamics will be modest.
 
  Given that precise and practical alternatives are now available, we feel that there is no justification for the use of biased graph randomization 
  processes. 
In those cases where we seek to generate unbiased random directed graphs with in- and out-degrees identical to some observed network, our canonical MCMC process  would take  the observed graph as its seed and take care of the required unbiased sampling. In those cases where we specify degree sequences {\em ab initio}, without having a seed graph, one may use the Molloy-Reed algorithm to generate a (biased) seed graph prior to running our algorithm. 

In addition to being rigorously free of entropic sampling bias, our present canonical MCMC process is also able to generate directed degree-constrained networks with any arbitrary specified sampling probabilities. We have shown examples of the generation of synthetic graphs generated with precisely controlled expectation values for the degree-degree correlation kernels, where the imposed sampling measure is a maximum  entropy distribution on the set of graphs with prescribed degrees, with degree correlations imposed as a soft constraint. Degree correlation is a promising candidate to define a better null model, as it has been observed in the literature to act as a `signature' distinguishing different types of networks (e.g. \cite{Maslov04, Newman02} ).

 Two directions for future research could be to look at weighted networks (e.g. to integrate our ideas with those in papers such as \cite{Ansmann11}), or at bipartite networks (which also have interesting applications, see e.g. \cite{Basler11}). Furthermore, it would seem appropriate in the field of network hypothesis testing to take more seriously the nontrivial number of short loops in biological signalling systems. Whenever we randomize within the large amorphous space of graphs that inherit from the biological network only the degree sequence, we are effectively running a dynamics on graphs that are locally tree-like,  where (conveniently) the mobility issues are minor. But we know already that this large set will typically produce null models that are very much unlike biological networks, for that same reason. How informative are small p-values in this context?
  
\section*{Acknowledgements}

 This study was supported by the Biotechnology and Biological Sciences Research Council of the United Kingdom. 
 It is our pleasure to thank Thomas Schlitt for providing gene regulation network data.

\bibliography{randomising_directed_networks}

\begin{thebibliography}{10}%
\makeatletter
\providecommand \@ifxundefined [1]{%
 \ifx #1\undefined \expandafter \@firstoftwo
 \else \expandafter \@secondoftwo
\fi
}%
\providecommand \@ifnum [1]{%
 \ifnum #1\expandafter \@firstoftwo
 \else \expandafter \@secondoftwo
\fi
}%
\providecommand \enquote [1]{``#1''}%
\providecommand \bibnamefont  [1]{#1}%
\providecommand \bibfnamefont [1]{#1}%
\providecommand \citenamefont [1]{#1}%
\providecommand\href[0]{\@sanitize\@href}%
\providecommand\@href[1]{\endgroup\@@startlink{#1}\endgroup\@@href}%
\providecommand\@@href[1]{#1\@@endlink}%
\providecommand \@sanitize [0]{\begingroup\catcode`\&12\catcode`\#12\relax}%
\@ifxundefined \pdfoutput {\@firstoftwo}{%
 \@ifnum{\z@=\pdfoutput}{\@firstoftwo}{\@secondoftwo}%
}{%
 \providecommand\@@startlink[1]{\leavevmode\special{html:<a href="#1">}}%
 \providecommand\@@endlink[0]{\special{html:</a>}}%
}{%
 \providecommand\@@startlink[1]{%
  \leavevmode
  \pdfstartlink
   attr{/Border[0 0 1 ]/H/I/C[0 1 1]}%
   user{/Subtype/Link/A<</Type/Action/S/URI/URI(#1)>>}%
  \relax
 }%
 \providecommand\@@endlink[0]{\pdfendlink}%
}%
\providecommand \url  [0]{\begingroup\@sanitize \@url }%
\providecommand \@url [1]{\endgroup\@href {#1}{\urlprefix}}%
\providecommand \urlprefix [0]{URL }%
\providecommand \Eprint[0]{\href }%
\@ifxundefined \urlstyle {%
  \providecommand \doi [1]{doi:\discretionary{}{}{}#1}%
}{%
  \providecommand \doi [0]{doi:\discretionary{}{}{}\begingroup
  \urlstyle{rm}\Url }%
}%
\providecommand \doibase [0]{http://dx.doi.org/}%
\providecommand \Doi[1]{\href{\doibase#1}}%
\providecommand \bibAnnote [3]{%
  \BibitemShut{#1}%
  \begin{quotation}\noindent
    \textsc{Key:}\ #2\\\textsc{Annotation:}\ #3%
  \end{quotation}%
}%
\providecommand \bibAnnoteFile [2]{%
  \IfFileExists{#2}{\bibAnnote {#1} {#2} {\input{#2}}}{}%
}%
\providecommand \typeout [0]{\immediate \write \m@ne }%
\providecommand \selectlanguage [0]{\@gobble}%
\providecommand \bibinfo [0]{\@secondoftwo}%
\providecommand \bibfield [0]{\@secondoftwo}%
\providecommand \translation [1]{[#1]}%
\providecommand \BibitemOpen[0]{}%
\providecommand \bibitemStop [0]{}%
\providecommand \bibitemNoStop [0]{.\EOS\space}%
\providecommand \EOS [0]{\spacefactor3000\relax}%
\providecommand \BibitemShut [1]{\csname bibitem#1\endcsname}%
\bibitem{MolloyReed}%
  \BibitemOpen
  \bibfield{author}{%
  \bibinfo {author} {\bibfnamefont{M.}~\bibnamefont{Molloy}}\ and\ \bibinfo
  {author} {\bibfnamefont{B.}~\bibnamefont{Reed}},\ }%
  \bibfield{journal}{%
  \bibinfo {journal} {Random Structures \& Algorithms}\ }%
  \textbf{\bibinfo {volume} {6}},\ \bibinfo {pages} {161} (\bibinfo {year}
  {1995}),\
  \url{http://citeseerx.ist.psu.edu/viewdoc/summary?doi=10.1.1.24.6195}%
  \bibAnnoteFile{NoStop}{MolloyReed}%
\bibitem{Barabasi-Albert}%
  \BibitemOpen
  \bibfield{author}{%
  \bibinfo {author} {\bibfnamefont{R.}~\bibnamefont{Albert}}\ and\ \bibinfo
  {author} {\bibfnamefont{A.~L.}\ \bibnamefont{Barab\'{a}si}},\ }%
  \bibfield{journal}{%
  \Doi{10.1103/RevModPhys.74.47}{\bibinfo {journal} {Reviews of Modern
  Physics}}\ }%
  \textbf{\bibinfo {volume} {74}},\ \bibinfo {pages} {47} (\bibinfo {month}
  {Jan.}\ \bibinfo {year} {2002}),\
  \url{http://dx.doi.org/10.1103/RevModPhys.74.47}%
  \bibAnnoteFile{NoStop}{Barabasi-Albert}%
\bibitem{Rao96}%
  \BibitemOpen
  \bibfield{author}{%
  \bibinfo {author} {\bibfnamefont{A.~R.}\ \bibnamefont{Rao}}, \bibinfo
  {author} {\bibfnamefont{R.}~\bibnamefont{Jana}},\ and\ \bibinfo {author}
  {\bibfnamefont{S.}~\bibnamefont{Bandyopadhyay}},\ }%
  \bibfield{journal}{%
  \bibinfo {journal} {Sankhyā: The Indian Journal of Statistics, Series A}\ }%
  \textbf{\bibinfo {volume} {58}} (\bibinfo {year} {1996}),\ ISSN \bibinfo
  {issn} {0581572X},\ \doi{\bibinfo {doi} {10.2307/25051102}},\
  \url{http://dx.doi.org/10.2307/25051102}%
  \bibAnnoteFile{NoStop}{Rao96}%
\bibitem{Squartini11_theory}%
  \BibitemOpen
  \bibfield{author}{%
  \bibinfo {author} {\bibfnamefont{T.}~\bibnamefont{Squartini}}, \bibinfo
  {author} {\bibfnamefont{G.}~\bibnamefont{Fagiolo}},\ and\ \bibinfo {author}
  {\bibfnamefont{D.}~\bibnamefont{Garlaschelli}},\ }%
  \bibfield{journal}{%
  \Doi{10.1103/PhysRevE.84.046118}{\bibinfo {journal} {Phys. Rev. E}}\ }%
  \textbf{\bibinfo {volume} {84}},\ \bibinfo {pages} {046118} (\bibinfo {month}
  {Oct}\ \bibinfo {year} {2011}),\
  \url{http://link.aps.org/doi/10.1103/PhysRevE.84.046118}%
  \bibAnnoteFile{NoStop}{Squartini11_theory}%
\bibitem{Squartini11_application}%
  \BibitemOpen
  \bibfield{author}{%
  \bibinfo {author} {\bibfnamefont{T.}~\bibnamefont{Squartini}}, \bibinfo
  {author} {\bibfnamefont{G.}~\bibnamefont{Fagiolo}},\ and\ \bibinfo {author}
  {\bibfnamefont{D.}~\bibnamefont{Garlaschelli}}}%
   (\bibinfo {year} {2011}),\ \url{http://arxiv.org/abs/1103.1243}%
  \bibAnnoteFile{NoStop}{Squartini11_application}%
\bibitem{shen-orr02}%
  \BibitemOpen
  \bibfield{author}{%
  \bibinfo {author} {\bibfnamefont{S.~S.}\ \bibnamefont{Shen-Orr}}, \bibinfo
  {author} {\bibfnamefont{R.}~\bibnamefont{Milo}}, \bibinfo {author}
  {\bibfnamefont{S.}~\bibnamefont{Mangan}},\ and\ \bibinfo {author}
  {\bibfnamefont{U.}~\bibnamefont{Alon}},\ }%
  \bibfield{journal}{%
  \Doi{10.1038/ng881}{\bibinfo {journal} {Nature Genetics}}\ }%
  \textbf{\bibinfo {volume} {31}},\ \bibinfo {pages} {64} (\bibinfo {month}
  {Apr.}\ \bibinfo {year} {2002}),\ ISSN \bibinfo {issn} {1061-4036},\
  \url{http://dx.doi.org/10.1038/ng881}%
  \bibAnnoteFile{NoStop}{shen-orr02}%
\bibitem{Power}%
  \BibitemOpen
  \bibfield{author}{%
  \bibinfo {author} {\bibfnamefont{G.~A.}\ \bibnamefont{Pagani}}\ and\ \bibinfo
  {author} {\bibfnamefont{M.}~\bibnamefont{Aiello}},\ }%
  \enquote{\bibinfo {title} {The power grid as a complex network: a survey},}\
  (\bibinfo {month} {May}\ \bibinfo {year} {2011}),\
  \url{http://arxiv.org/abs/1105.3338}%
  \bibAnnoteFile{NoStop}{Power}%
\bibitem{Japan_interfirm}%
  \BibitemOpen
  \bibfield{author}{%
  \bibinfo {author} {\bibfnamefont{T.}~\bibnamefont{Ohnishi}}, \bibinfo
  {author} {\bibfnamefont{H.}~\bibnamefont{Takayasu}},\ and\ \bibinfo {author}
  {\bibfnamefont{M.}~\bibnamefont{Takayasu}},\ }%
  \bibfield{journal}{%
  \bibinfo {journal} {Journal of Economic Interaction and Coordination}}%
   (\bibinfo {month} {Jun.}\ \bibinfo {year} {2010}),\ ISSN \bibinfo {issn}
  {1860-711X},\ \doi{\bibinfo {doi} {10.1007/s11403-010-0066-6}},\
  \url{http://dx.doi.org/10.1007/s11403-010-0066-6}%
  \bibAnnoteFile{NoStop}{Japan_interfirm}%
\bibitem{newman_social}%
  \BibitemOpen
  \bibfield{author}{%
  \bibinfo {author} {\bibfnamefont{M.~E.~J.}\ \bibnamefont{Newman}}, \bibinfo
  {author} {\bibfnamefont{D.~J.}\ \bibnamefont{Watts}},\ and\ \bibinfo {author}
  {\bibfnamefont{S.~H.}\ \bibnamefont{Strogatz}},\ }%
  \bibfield{journal}{%
  \Doi{10.1073/pnas.012582999}{\bibinfo {journal} {Proceedings of the National
  Academy of Sciences of the United States of America}}\ }%
  \textbf{\bibinfo {volume} {99}},\ \bibinfo {pages} {2566} (\bibinfo {month}
  {Feb.}\ \bibinfo {year} {2002}),\ ISSN \bibinfo {issn} {0027-8424},\
  \url{http://dx.doi.org/10.1073/pnas.012582999}%
  \bibAnnoteFile{NoStop}{newman_social}%
\bibitem{Maslov02}%
  \BibitemOpen
  \bibfield{author}{%
  \bibinfo {author} {\bibfnamefont{S.}~\bibnamefont{Maslov}}\ and\ \bibinfo
  {author} {\bibfnamefont{K.}~\bibnamefont{Sneppen}},\ }%
  \bibfield{journal}{%
  \Doi{10.1126/science.1065103}{\bibinfo {journal} {Science}}\ }%
  \textbf{\bibinfo {volume} {296}},\ \bibinfo {pages} {910} (\bibinfo {month}
  {May}\ \bibinfo {year} {2002}),\ ISSN \bibinfo {issn} {1095-9203},\
  \url{http://dx.doi.org/10.1126/science.1065103}%
  \bibAnnoteFile{NoStop}{Maslov02}%
\bibitem{Ecology}%
  \BibitemOpen
  \bibfield{author}{%
  \bibinfo {author} {\bibfnamefont{N.~J.}\ \bibnamefont{Gotelli}}\ and\
  \bibinfo {author} {\bibfnamefont{W.}~\bibnamefont{Ulrich}},\ }%
  \bibfield{journal}{%
  \Doi{10.1111/j.1600-0706.2011.20301.x}{\bibinfo {journal} {Oikos}},\ \bibinfo
  {pages} {no}}%
   (\bibinfo {year} {2011}),\ ISSN \bibinfo {issn} {1600-0706},\
  \url{http://dx.doi.org/10.1111/j.1600-0706.2011.20301.x}%
  \bibAnnoteFile{NoStop}{Ecology}%
\bibitem{King04}%
  \BibitemOpen
  \bibfield{author}{%
  \bibinfo {author} {\bibfnamefont{O.~D.}\ \bibnamefont{King}},\ }%
  \bibfield{journal}{%
  \bibinfo {journal} {Phys. Rev. E}\ }%
  \textbf{\bibinfo {volume} {70}},\ \bibinfo {pages} {058101+} (\bibinfo {year}
  {2004})%
  \bibAnnoteFile{NoStop}{King04}%
\bibitem{Klein-Hennig}%
  \BibitemOpen
  \bibfield{author}{%
  \bibinfo {author} {\bibfnamefont{H.}~\bibnamefont{Klein-Hennig}}\ and\
  \bibinfo {author} {\bibfnamefont{A.~K.}\ \bibnamefont{Hartmann}},\ }%
  \enquote{\bibinfo {title} {Bias in generation of random graphs},}\  (\bibinfo
  {year} {2011}),\ \url{http://arxiv.org/abs/1107.5734}%
  \bibAnnoteFile{NoStop}{Klein-Hennig}%
\bibitem{citeulike:9790164}%
  \BibitemOpen
  \bibfield{author}{%
  \bibinfo {author} {\bibfnamefont{H.}~\bibnamefont{Kim}}, \bibinfo {author}
  {\bibfnamefont{C.~I.}\ \bibnamefont{Del~Genio}}, \bibinfo {author}
  {\bibfnamefont{K.~E.}\ \bibnamefont{Bassler}},\ and\ \bibinfo {author}
  {\bibfnamefont{Z.}~\bibnamefont{Toroczkai}}}%
   (\bibinfo {month} {Sep.}\ \bibinfo {year} {2011}),\
  \Eprint{http://arxiv.org/abs/1109.4590}{arXiv:1109.4590},\
  \url{http://arxiv.org/abs/1109.4590}%
  \bibAnnoteFile{NoStop}{citeulike:9790164}%
\bibitem{Itz2003}%
  \BibitemOpen
  \bibfield{author}{%
  \bibinfo {author} {\bibfnamefont{S.}~\bibnamefont{Itzkovitz}}, \bibinfo
  {author} {\bibfnamefont{R.}~\bibnamefont{Milo}}, \bibinfo {author}
  {\bibfnamefont{N.}~\bibnamefont{Kashtan}}, \bibinfo {author}
  {\bibfnamefont{G.}~\bibnamefont{Ziv}},\ and\ \bibinfo {author}
  {\bibfnamefont{U.}~\bibnamefont{Alon}},\ }%
  \bibfield{journal}{%
  \Doi{10.1103/PhysRevE.68.026127}{\bibinfo {journal} {Physical Review E}}\ }%
  \textbf{\bibinfo {volume} {68}},\ \bibinfo {pages} {026127+} (\bibinfo
  {month} {Aug.}\ \bibinfo {year} {2003}),\
  \url{http://dx.doi.org/10.1103/PhysRevE.68.026127}%
  \bibAnnoteFile{NoStop}{Itz2003}%
\bibitem{Coolen09}%
  \BibitemOpen
  \bibfield{author}{%
  \bibinfo {author} {\bibfnamefont{A.~C.~C.}\ \bibnamefont{Coolen}}, \bibinfo
  {author} {\bibfnamefont{A.}~\bibnamefont{De~Martino}},\ and\ \bibinfo
  {author} {\bibfnamefont{A.}~\bibnamefont{Annibale}},\ }%
  \bibfield{journal}{%
  \bibinfo {journal} {J. Stat. Phys.}}%
   (\bibinfo {month} {May}\ \bibinfo {year} {2009}),\
  \Eprint{http://arxiv.org/abs/0905.4155}{arXiv:0905.4155}%
  \bibAnnoteFile{NoStop}{Coolen09}%
\bibitem{Milo04}%
  \BibitemOpen
  \bibfield{author}{%
  \bibinfo {author} {\bibfnamefont{R.}~\bibnamefont{Milo}}, \bibinfo {author}
  {\bibfnamefont{N.}~\bibnamefont{Kashtan}}, \bibinfo {author}
  {\bibfnamefont{S.}~\bibnamefont{Itzkovitz}}, \bibinfo {author}
  {\bibfnamefont{M.~E.~J.}\ \bibnamefont{Newman}},\ and\ \bibinfo {author}
  {\bibfnamefont{U.}~\bibnamefont{Alon}},\ }%
  \enquote{\bibinfo {title} {{On the uniform generation of random graphs with
  prescribed degree sequences}},}\  (\bibinfo {month} {May}\ \bibinfo {year}
  {2004}),\
  \Eprint{http://arxiv.org/abs/cond-mat/0312028}{arXiv:cond-mat/0312028},\
  \url{http://arxiv.org/abs/cond-mat/0312028}%
  \bibAnnoteFile{NoStop}{Milo04}%
\bibitem{Seidel}%
  \BibitemOpen
  \bibfield{author}{%
  \bibinfo {author} {\bibfnamefont{J.~J.}\ \bibnamefont{Seidel}},\ }%
  in\ \emph{\bibinfo {booktitle} {In Colloquio Internazionale sulle Teorie
  Combinatorie Tomo I}},\ \bibinfo {editor} {edited by\ \bibinfo {editor}
  {\bibfnamefont{A.}~\bibnamefont{Doe}}}\ (\bibinfo {year} {1973})\ pp.\
  \bibinfo {pages} {481--511}%
  \bibAnnoteFile{NoStop}{Seidel}%
\bibitem{Taylor}%
  \BibitemOpen
  \bibfield{author}{%
  \bibinfo {author} {\bibfnamefont{R.}~\bibnamefont{Taylor}},\ }%
  \emph{\bibinfo {title} {Combinatorial Mathematics VIII}}\ (\bibinfo
  {publisher} {Springer},\ \bibinfo {year} {1981})%
  \bibAnnoteFile{NoStop}{Taylor}%
\bibitem{Roberts11}%
  \BibitemOpen
  \bibfield{author}{%
  \bibinfo {author} {\bibfnamefont{E.~S.}\ \bibnamefont{Roberts}}, \bibinfo
  {author} {\bibfnamefont{T.}~\bibnamefont{Schlitt}},\ and\ \bibinfo {author}
  {\bibfnamefont{A.~C.~C.}\ \bibnamefont{Coolen}},\ }%
  \bibfield{journal}{%
  \bibinfo {journal} {J. Phys. A}\ }%
  \textbf{\bibinfo {volume} {44}},\ \bibinfo {pages} {275002} (\bibinfo {year}
  {2011}),\ \url{http://stacks.iop.org/1751-8121/44/i=27/a=275002}%
  \bibAnnoteFile{NoStop}{Roberts11}%
\bibitem{MersenneTwister}%
  \BibitemOpen
  \bibfield{author}{%
  \bibinfo {author} {\bibfnamefont{M.}~\bibnamefont{Matsumoto}}\ and\ \bibinfo
  {author} {\bibfnamefont{T.}~\bibnamefont{Nishimura}},\ }%
  \bibfield{journal}{%
  \Doi{10.1145/272991.272995}{\bibinfo {journal} {ACM Trans. Model. Comput.
  Simul.}}\ }%
  \textbf{\bibinfo {volume} {8}},\ \bibinfo {pages} {3} (\bibinfo {month}
  {Jan.}\ \bibinfo {year} {1998}),\ ISSN \bibinfo {issn} {1049-3301},\
  \url{http://dx.doi.org/10.1145/272991.272995}%
  \bibAnnoteFile{NoStop}{MersenneTwister}%
\bibitem{Hughes}%
  \BibitemOpen
  \bibfield{author}{%
  \bibinfo {author} {\bibfnamefont{T.~R.}\ \bibnamefont{Hughes}}, \bibinfo
  {author} {\bibfnamefont{M.~J.}\ \bibnamefont{Marton}}, \bibinfo {author}
  {\bibfnamefont{A.~R.}\ \bibnamefont{Jones}}, \bibinfo {author}
  {\bibfnamefont{C.~J.}\ \bibnamefont{Roberts}}, \bibinfo {author}
  {\bibfnamefont{R.}~\bibnamefont{Stoughton}}, \bibinfo {author}
  {\bibfnamefont{C.~D.}\ \bibnamefont{Armour}}, \bibinfo {author}
  {\bibfnamefont{H.~A.}\ \bibnamefont{Bennett}}, \bibinfo {author}
  {\bibfnamefont{E.}~\bibnamefont{Coffey}}, \bibinfo {author}
  {\bibfnamefont{H.}~\bibnamefont{Dai}}, \bibinfo {author}
  {\bibfnamefont{Y.~D.}\ \bibnamefont{He}}, \bibinfo {author}
  {\bibfnamefont{M.~J.}\ \bibnamefont{Kidd}}, \bibinfo {author}
  {\bibfnamefont{A.~M.}\ \bibnamefont{King}}, \bibinfo {author}
  {\bibfnamefont{M.~R.}\ \bibnamefont{Meyer}}, \bibinfo {author}
  {\bibfnamefont{D.}~\bibnamefont{Slade}}, \bibinfo {author}
  {\bibfnamefont{P.~Y.}\ \bibnamefont{Lum}}, \bibinfo {author}
  {\bibfnamefont{S.~B.}\ \bibnamefont{Stepaniants}}, \bibinfo {author}
  {\bibfnamefont{D.~D.}\ \bibnamefont{Shoemaker}}, \bibinfo {author}
  {\bibfnamefont{D.}~\bibnamefont{Gachotte}}, \bibinfo {author}
  {\bibfnamefont{K.}~\bibnamefont{Chakraburtty}}, \bibinfo {author}
  {\bibfnamefont{J.}~\bibnamefont{Simon}}, \bibinfo {author}
  {\bibfnamefont{M.}~\bibnamefont{Bard}},\ and\ \bibinfo {author}
  {\bibfnamefont{S.~H.}\ \bibnamefont{Friend}},\ }%
  \bibfield{journal}{%
  \bibinfo {journal} {Cell}\ }%
  \textbf{\bibinfo {volume} {102}},\ \bibinfo {pages} {109} (\bibinfo {month}
  {Jul.}\ \bibinfo {year} {2000}),\ ISSN \bibinfo {issn} {0092-8674},\
  \url{http://view.ncbi.nlm.nih.gov/pubmed/10929718}%
  \bibAnnoteFile{NoStop}{Hughes}%
\bibitem{Harbison}%
  \BibitemOpen
  \bibfield{author}{%
  \bibinfo {author} {\bibfnamefont{C.~T.}\ \bibnamefont{Harbison}}, \bibinfo
  {author} {\bibfnamefont{D.~B.}\ \bibnamefont{Gordon}}, \bibinfo {author}
  {\bibfnamefont{T.~I.~I.}\ \bibnamefont{Lee}}, \bibinfo {author}
  {\bibfnamefont{N.~J.}\ \bibnamefont{Rinaldi}}, \bibinfo {author}
  {\bibfnamefont{K.~D.}\ \bibnamefont{Macisaac}}, \bibinfo {author}
  {\bibfnamefont{T.~W.}\ \bibnamefont{Danford}}, \bibinfo {author}
  {\bibfnamefont{N.~M.}\ \bibnamefont{Hannett}}, \bibinfo {author}
  {\bibfnamefont{J.-B.~B.}\ \bibnamefont{Tagne}}, \bibinfo {author}
  {\bibfnamefont{D.~B.}\ \bibnamefont{Reynolds}}, \bibinfo {author}
  {\bibfnamefont{J.}~\bibnamefont{Yoo}}, \bibinfo {author}
  {\bibfnamefont{E.~G.}\ \bibnamefont{Jennings}}, \bibinfo {author}
  {\bibfnamefont{J.}~\bibnamefont{Zeitlinger}}, \bibinfo {author}
  {\bibfnamefont{D.~K.}\ \bibnamefont{Pokholok}}, \bibinfo {author}
  {\bibfnamefont{M.}~\bibnamefont{Kellis}}, \bibinfo {author}
  {\bibfnamefont{P.~A.}\ \bibnamefont{Rolfe}}, \bibinfo {author}
  {\bibfnamefont{K.~T.}\ \bibnamefont{Takusagawa}}, \bibinfo {author}
  {\bibfnamefont{E.~S.}\ \bibnamefont{Lander}}, \bibinfo {author}
  {\bibfnamefont{D.~K.}\ \bibnamefont{Gifford}}, \bibinfo {author}
  {\bibfnamefont{E.}~\bibnamefont{Fraenkel}},\ and\ \bibinfo {author}
  {\bibfnamefont{R.~A.}\ \bibnamefont{Young}},\ }%
  \bibfield{journal}{%
  \Doi{10.1038/nature02800}{\bibinfo {journal} {Nature}}\ }%
  \textbf{\bibinfo {volume} {431}},\ \bibinfo {pages} {99} (\bibinfo {month}
  {Sep.}\ \bibinfo {year} {2004}),\ ISSN \bibinfo {issn} {1476-4687},\
  \url{http://dx.doi.org/10.1038/nature02800}%
  \bibAnnoteFile{NoStop}{Harbison}%
\bibitem{Fernandes10}%
  \BibitemOpen
  \bibfield{author}{%
  \bibinfo {author} {\bibfnamefont{L.~P.}\ \bibnamefont{Fernandes}}, \bibinfo
  {author} {\bibfnamefont{A.}~\bibnamefont{Annibale}}, \bibinfo {author}
  {\bibfnamefont{J.}~\bibnamefont{Kleinjung}}, \bibinfo {author}
  {\bibfnamefont{A.~C.~C.}\ \bibnamefont{Coolen}},\ and\ \bibinfo {author}
  {\bibfnamefont{F.}~\bibnamefont{Fraternali}},\ }%
  \bibfield{journal}{%
  \Doi{10.1371/journal.pone.0012083}{\bibinfo {journal} {PLoS ONE}}\ }%
  \textbf{\bibinfo {volume} {5}},\ \bibinfo {pages} {e12083+} (\bibinfo {month}
  {Aug.}\ \bibinfo {year} {2010}),\ ISSN \bibinfo {issn} {1932-6203},\
  \url{http://dx.doi.org/10.1371/journal.pone.0012083}%
  \bibAnnoteFile{NoStop}{Fernandes10}%
\bibitem{Artzy-Randrup}%
  \BibitemOpen
  \bibfield{author}{%
  \bibinfo {author} {\bibfnamefont{Y.}~\bibnamefont{Artzy-Randrup}}\ and\
  \bibinfo {author} {\bibfnamefont{L.}~\bibnamefont{Stone}},\ }%
  \bibfield{journal}{%
  \Doi{10.1103/PhysRevE.72.056708}{\bibinfo {journal} {Phys. Rev. E}}\ }%
  \textbf{\bibinfo {volume} {72}},\ \bibinfo {pages} {056708} (\bibinfo {month}
  {Nov}\ \bibinfo {year} {2005}),\
  \url{http://link.aps.org/doi/10.1103/PhysRevE.72.056708}%
  \bibAnnoteFile{NoStop}{Artzy-Randrup}%
\bibitem{Maslov04}%
  \BibitemOpen
  \bibfield{author}{%
  \bibinfo {author} {\bibfnamefont{S.}~\bibnamefont{Maslov}}, \bibinfo {author}
  {\bibfnamefont{K.}~\bibnamefont{Sneppen}},\ and\ \bibinfo {author}
  {\bibfnamefont{A.}~\bibnamefont{Zaliznyak}},\ }%
  \bibfield{journal}{%
  \Doi{10.1016/j.physa.2003.06.002}{\bibinfo {journal} {Physica A: Statistical
  Mechanics and its Applications}}\ }%
  \textbf{\bibinfo {volume} {333}},\ \bibinfo {pages} {529 } (\bibinfo {year}
  {2004}),\ ISSN \bibinfo {issn} {0378-4371},\
  \url{http://www.sciencedirect.com/science/article/pii/S0378437103008409}%
  \bibAnnoteFile{NoStop}{Maslov04}%
\bibitem{Newman02}%
  \BibitemOpen
  \bibfield{author}{%
  \bibinfo {author} {\bibfnamefont{M.~E.~J.}\ \bibnamefont{Newman}},\ }%
  \bibfield{journal}{%
  \Doi{10.1103/PhysRevLett.89.208701}{\bibinfo {journal} {Phys. Rev. Lett.}}\
  }%
  \textbf{\bibinfo {volume} {89}},\ \bibinfo {pages} {208701} (\bibinfo {month}
  {Oct}\ \bibinfo {year} {2002}),\
  \url{http://link.aps.org/doi/10.1103/PhysRevLett.89.208701}%
  \bibAnnoteFile{NoStop}{Newman02}%
\bibitem{Ansmann11}%
  \BibitemOpen
  \bibfield{author}{%
  \bibinfo {author} {\bibfnamefont{G.}~\bibnamefont{Ansmann}}\ and\ \bibinfo
  {author} {\bibfnamefont{K.}~\bibnamefont{Lehnertz}},\ }%
  \bibfield{journal}{%
  \Doi{10.1103/PhysRevE.84.026103}{\bibinfo {journal} {Phys. Rev. E}}\ }%
  \textbf{\bibinfo {volume} {84}},\ \bibinfo {pages} {026103+} (\bibinfo
  {month} {Aug.}\ \bibinfo {year} {2011}),\
  \url{http://dx.doi.org/10.1103/PhysRevE.84.026103}%
  \bibAnnoteFile{NoStop}{Ansmann11}%
\bibitem{Basler11}%
  \BibitemOpen
  \bibfield{author}{%
  \bibinfo {author} {\bibfnamefont{G.}~\bibnamefont{Basler}}, \bibinfo {author}
  {\bibfnamefont{O.}~\bibnamefont{Ebenh\"{o}h}}, \bibinfo {author}
  {\bibfnamefont{J.}~\bibnamefont{Selbig}},\ and\ \bibinfo {author}
  {\bibfnamefont{Z.}~\bibnamefont{Nikoloski}},\ }%
  \bibfield{journal}{%
  \bibinfo {journal} {Bioinformatics}}%
   (\bibinfo {month} {Mar.}\ \bibinfo {year} {2011}),\ ISSN \bibinfo {issn}
  {1367-4811},\ \doi{\bibinfo {doi} {10.1093/bioinformatics/btr145}},\
  \url{http://dx.doi.org/10.1093/bioinformatics/btr145}%
  \bibAnnoteFile{NoStop}{Basler11}%
\end{thebibliography}%
\bibliographystyle{apsrev4-1}

\appendix

\section{Efficient calculation of changes in mobility terms following one move}

\label{app:mobility_change}

\noindent
Calculating the mobility $n(\bc)$ terms is computationally heavy. Given that our moves are simple and standard, we follow the alternative route in \cite{Coolen09}  and derive formulae for calculating the {\em change} in mobility due to one move, so that we can avoid  repeated heavy matrix multiplications at each time step.  

\subsection{Change in $n_{\square}(\bc)$ following one square-type move}

\noindent
Without loss of generality, define our square move to be the transformation between matrix $c$ and $x$, involving four nodes $(a,b,c,d)$, such that 
for all $(i,j)$: $x_{ij} = c_{ij}+\Delta_{ij}$, with
\begin{eqnarray*}
\label{square_move}
\Delta_{ij}&=&  \delta_{ia}\delta_{jd} + \delta_{ic}\delta_{jb} - \delta_{ia}\delta_{jb} - \delta_{ic}\delta_{jd}
\end{eqnarray*}
We now determine the overall change induced in $n_{\square}(\bc)$ by finding the impact of an edge swap 
on each term in  (\ref{eq:square_mobility_term}). 
on the right hand side of the expression above. 
\begin{itemize}
\item Term 1:
\begin{eqnarray*}
&&
  {\rm Tr}(\bx \bx^\dag \bx \bx^\dag)  - {\rm Tr}(\bc \bc^\dag \bc \bc^\dag)= \sum_{ijkm}\Big[c_{ij}c_{kj}c_{km}c_{im}  
 \\
 && \hspace*{10mm}
 -(c_{ij} + \Delta_{ij})(c_{kj} + \Delta_{kj})
 		(c_{km} + \Delta_{km})(c_{im} + \Delta_{im})\Big]
\\
&  &~~~= \Delta_{ij}c_{kj}c_{km}c_{im} + ...  + \Delta_{ij}\Delta_{kj}c_{km}c_{im}  + ... \\
 & &~~~~~~+ \Delta_{ij}\Delta_{kj}\Delta_{km}c_{im} + ...+ \Delta_{ij}\Delta_{kj}\Delta_{km}\Delta_{im} 
\end{eqnarray*}
where $...$ refers in each case to three similar terms (with their appropriate indices).
Let us inspect what happens when two $\Delta$ terms are multiplied together. We might have the first suffix repeated, 
the second suffix repeated, or no repeated sufficies: 
\begin{eqnarray}
&&
\Delta_{ij} \Delta_{im} = 2[ \delta_{jd}(\delta_{md} - \delta_{mb}) + \delta_{jb}(\delta_{mb} - \delta_{md}) ] 
\nonumber \\
&&
\Delta_{ij} \Delta_{kj} = 2[ \delta_{ia}(\delta_{ka} - \delta_{kc}) + \delta_{ic}(\delta_{kc} - \delta_{ka}) ]
\label{double_delta}
\end{eqnarray}
One immediately observes that 
\begin{eqnarray*}
 &&\sum_{ijkm}\Delta_{ij}\Delta_{kj}\Delta_{km}\Delta_{im} \\
 &&\hspace*{3mm}
  =4  \sum_{ik}\left[ \delta_{ia}\delta_{ia}(\delta_{ka} \delta_{ka}+ \delta_{kc}\delta_{kc}) + \delta_{ic}\delta_{ic}(\delta_{kc} \delta_{kc} + \delta_{ka}\delta_{ka})\right] \hspace*{-10mm}\\
  &&\hspace*{3mm}= 16
\end{eqnarray*}
To handle two $\Delta$ terms with different sufficies we use
\begin{eqnarray}
\Delta_{ij} c_{kj} = c_{kb} \left( \delta_{ic} - \delta_{ia}\right) + c_{kd} \left( \delta_{ia} - \delta_{ic}\right) 
\label{delta_and_c}
\end{eqnarray}
which leads us to
\begin{eqnarray*}
\sum_{ijkm}\Delta_{ij} c_{kj}\Delta_{km} c_{im}  = 4
\end{eqnarray*}
Returning to the result \ref{double_delta} it follows that 
\begin{eqnarray*}
\Delta_{ij} \Delta_{kj} c_{im} c_{km} &=& 2\left( \delta_{ia}(\delta_{ka} - \delta_{kc}) + \delta_{ic}(\delta_{kc} - \delta_{ka}) \right) c_{im} c_{km}
\hspace*{-10mm} \\ \nonumber
&=& 2 (k_a^{\rm out} + k_c^{\rm out}) - 4 c_{am}c_{cm}
\end{eqnarray*}
and the symmetric term gives
\begin{eqnarray*}
\Delta_{ij} \Delta_{im} c_{kj} c_{km}&= 2 (k_d^{\rm in} + k_b^{\rm in}) - 4 c_{id}c_{ib}
\end{eqnarray*}
For the third order terms we  combine (\ref{double_delta}) and (\ref{delta_and_c}):
\begin{eqnarray*}
&&\hspace*{-3mm} 
\sum_{ijkm}\Delta_{ij} \Delta_{im} \Delta_{kj} c_{km}= 2\sum_{ik}\Big[\left[ \delta_{ia}(\delta_{ka}\! -\! \delta_{kc}) + \delta_{ic}(\delta_{kc}\! -\! \delta_{ka}) \right] \hspace*{-10mm}
\\ \nonumber
										&&\hspace*{10mm} \times \left[ \delta_{ia}c_{kd} + \delta_{ic}c_{kb} - \delta_{ia}c_{kb} - \delta_{ic}c_{kd}\right]  \Big] \nonumber\\
	& & ~~~=2(c_{ad}-c_{cd}- c_{ab} +c_{cb}-c_{ab}+c_{cb}+c_{ad}-c_{cd})
	\hspace*{-10mm} \\
	&&~~~=	-8			
\end{eqnarray*}
By permutation of sufficies all such terms evaluate to $-8$. 
Finally we turn to the four terms where only one $\Delta$ appears,  corresponding to permutations of $\Delta_{ij}c_{kj}c_{km}c_{im}= c_{kd}c_{km}c_{am} + c_{kb}c_{km}c_{cm} - c_{kb}c_{km}c_{am} - c_{kd}c_{km}c_{cm}$.
Adding up all  separate elements above, we obtain the change in the square mobility term due to one application of a square move:
\begin{eqnarray}
&& \hspace*{-3mm}\Delta\Big[ \frac{1}{2} {\rm Tr}(\bc \bc^\dag \bc \bc^\dag)\Big]
= 
	2 (k_d^{\rm in} + k_b^{\rm in}+ k_a^{\rm out} + k_c^{\rm out}) 
	\nonumber
	\\
	&&+ 2 \left(c_{kd}c_{km}c_{am} + c_{kb}c_{km}c_{cm} - c_{kb}c_{km}c_{am} - c_{kd}c_{km}c_{cm} \right)\nonumber
	\hspace*{-10mm}
	\\
	&&
	- 4 \left(c_{id}c_{ib} + c_{am}c_{cm} +1\right)
\end{eqnarray}
\item Term 2:
\begin{eqnarray}&&
\Delta \Big[\sum_{ij}k^{\rm out}_i c_{ij}k^{\rm in}_j\Big] =   \sum_{ij}k^{\rm out}_i \left( x_{ij} - c_{ij} \right)k^{\rm in}_j 
\nonumber
\\
&&~~~~~~= \sum_{ij}k^{\rm out}_i \left[ \delta_{ia}\delta_{jd} + \delta_{ic}\delta_{jb} - \delta_{ia}\delta_{jb} - \delta_{ic}\delta_{jd}\right] k^{\rm in}_j  
\nonumber
\\
&& ~~~~~~=k_a^{\rm out}k_d^{\rm in} + k_c^{\rm out}k_b^{\rm in} - k_a^{\rm out}k_b^{\rm in} - k_c^{\rm out}k_d^{\rm in}
\end{eqnarray}
\item Term 3:
\begin{eqnarray*}
 &&\hspace*{-3mm}  
 {\rm Tr}(\mathbf{x} \mathbf{x}^\dag \mathbf{x}) - {\rm Tr}(\bc \bc^\dag \bc)  
 \\
 &&
 	=\sum_{ijk}\left[ c_{ij} + \Delta_{ij}\right] \left[ c_{kj} + \Delta_{kj}\right] \left[ c_{ki} + \Delta_{ki}\right] 	- c_{ij}c_{kj}c_{ki}  
 \\
 &&
  	 = \sum_{ijk}\Big[\Delta_{kj}c_{ij}c_{ki} +\Delta_{ij} c_{kj}c_{ki}  + \Delta_{ki}c_{ij}c_{kj}	+ \Delta_{ij}\Delta_{kj}c_{ki}
	 \\
&&	~~~~~~~~  + \Delta_{kj}\Delta_{ki}c_{ij}  
	 + \Delta_{ij}\Delta_{ki}c_{kj} + \Delta_{ij}\Delta_{kj}\Delta_{ki}\Big]
 \end{eqnarray*}
The product of two $\Delta$ terms gives
\begin{eqnarray*}
 \Delta_{ij}\Delta_{kj} 
 & =& \delta_{ik}\left( \delta_{jd}(\delta_{ia} - \delta_{ic}) + \delta_{jb}(\delta_{ic} - \delta_{ia})\right)
 \\
 &&
 - \delta_{jd}(\delta_{ia}\delta_{kc}+ \delta_{ic}\delta_{ka})
  - \delta_{jb}(\delta_{ic}\delta_{ka}+ \delta_{ia}\delta_{kc})
\end{eqnarray*}
but $
 \Delta_{ij}\Delta_{ki} 
 = 0
$,
and in a straightforward way we obtain
\begin{eqnarray*}
\sum_{ijk}
\Delta_{kj}c_{ij}c_{ki}&=&\sum_{ijk} \left[ \delta_{ka}\delta_{jd} + \delta_{kc}\delta_{jb} - \delta_{ka}\delta_{jb} - \delta_{kc}\delta_{jd}\right]c_{ij}c_{ki} 
\hspace*{-10mm}
\\
& =& \sum_i\Big[c_{ai} c_{id} + c_{ci} c_{ib} - c_{ai} c_{ib} - c_{ci} c_{id}\Big]
\end{eqnarray*}
Assembling all terms and their symmetric equivalents leads to an expression which can be summarised as 
\begin{eqnarray}
\Delta\Big[{\rm Tr}(\bc \bc^\dag \bc)\Big] &=& {\rm MutN}(a,d)+ {\rm MutN}(c,b) - {\rm MutN}(a,b)
\hspace*{-5mm}
\nonumber
\\
&& -{\rm MutN}(c,d) -2 (c_{bd}+ c_{db}+ c_{ac}+c_{ca})
\hspace*{-5mm}\nonumber
\\&&
\end{eqnarray}
\vspace*{-12mm} 

\noindent
where 
\begin{eqnarray}
{\rm MutN}(\alpha,\beta)=\sum_i \left[ c_{\alpha i}c_{i \beta} + c_{\alpha i}c_{\beta i} + c_{i \alpha }c_{ i \beta }\right] 
\end{eqnarray}
\item Term 5:
\begin{eqnarray}
&&\hspace*{-3mm} \Delta\left[ {\rm Tr}(\bc^2)\right] = {\rm Tr}(\bx^2)-{\rm Tr}(\bc^2)
\nonumber\\
&&= \sum_{ij}[c_{ij} + (\delta_{ia}\delta_{jd}+\delta_{ic}\delta_{jb}-\delta_{ia}\delta_{jb}-\delta_{ic}\delta_{jd})]
\nonumber\\
 &&~~\times [c_{ji} + (\delta_{ja}\delta_{id}+\delta_{jc}\delta_{ib}-\delta_{ja}\delta_{ib}-\delta_{jc}\delta_{id})]
 -{\rm Tr}(\bc^2)\hspace*{-10mm}
\nonumber \\
& &=  2(c_{da} + c_{bc} - c_{ba} - c_{dc})
\end{eqnarray}
\item Terms 4 and 6:\\[1mm]
The two terms $\frac{1}{2}N^2 \bra k\ket^2$ and 
$\sum_i k^{\rm out}_i k^{\rm in}_i$ do not change, since our stochastic process conserves all degrees. 
\end{itemize}
In combination, the above ingredients lead us to the following update formula for the square mobility (\ref{eq:square_mobility_term}), as a result of the 
edge swap (\ref{square_move}):
\begin{eqnarray}
\Delta n_{\square}&=& 
2 (k_d^{\rm in} + k_b^{\rm in}+ k_a^{\rm out} + k_c^{\rm out}) 
	\nonumber
	\\
	&&+ 2 \left(c_{kd}c_{km}c_{am} + c_{kb}c_{km}c_{cm} - c_{kb}c_{km}c_{am} - c_{kd}c_{km}c_{cm} \right)\nonumber
	\hspace*{-10mm}
	\\
	&&
	- 4 \left(c_{id}c_{ib} + c_{am}c_{cm} +1\right)\nonumber
	\\&&
-[k_a^{\rm out}k_d^{\rm in} + k_c^{\rm out}k_b^{\rm in} - k_a^{\rm out}k_b^{\rm in} - k_c^{\rm out}k_d^{\rm in}]
\nonumber
\\
&&
+ {\rm MutN}(a,d)+ {\rm MutN}(c,b) - {\rm MutN}(a,b)
\hspace*{-5mm}
\nonumber
\\
&& -{\rm MutN}(c,d) -2 (c_{bd}+ c_{db}+ c_{ac}+c_{ca})
\nonumber
\\
&&
+c_{da} + c_{bc} - c_{ba} - c_{dc}
\end{eqnarray}

\subsection{Change in $n_{\triangle}(\bc)$ following one square-type move}

\noindent
The different terms in the triangle mobility term (to be called Term 7, Term 8, Term 9 and Term 10, to avoid confusion with the previous section)
are
\begin{eqnarray*}
n(\bc)_{\triangle}
= & \frac{1}{3} 
	{\rm Tr}(\mathbf{c}^3) 
	- {\rm Tr}(\mathbf{c^{\updownarrow}} \mathbf{c}^2 ) 
	+ {\rm Tr}(\mathbf{c^{\updownarrow}}^2 \mathbf{c})
	- \frac{1}{3} {\rm Tr}(\mathbf{c^{\updownarrow}}^3)
	\\ \nonumber
\end{eqnarray*}
\begin{itemize}
\item 
Term 7:
\begin{eqnarray*}
\Delta {\rm Tr}(\mathbf{c}^3) &=& 
\sum_{ijk}\Big[x_{ij}x_{jk}x_{ki} -c_{ij}c_{jk}c_{ki}\Big]\nonumber\\
&=& 
 3 \sum_i\left( c_{ia}c_{di} + c_{ic}c_{bi}- c_{ia}c_{bi}- c_{ic}c_{di}\right) 
\end{eqnarray*}
\item Term 8:
\\[1mm]
Here we have to inspect first how the matrix $\bc^{\updownarrow}$ of double bonds is affected by a square move:
\begin{eqnarray*}
x^{\updownarrow}_{ij} = c^{\updownarrow}_{ij} + \Delta_{ij}^{\updownarrow} + \Delta_{ji}^{\updownarrow}
\end{eqnarray*}
with
\begin{eqnarray*}
 \Delta_{ij}^{\updownarrow} =\delta_{ia}\delta_{jd}c_{da} + \delta_{ic}\delta_{jb}c_{bc} - \delta_{ia}\delta_{jb}c_{ba} - \delta_{ic}\delta_{jd}c_{dc} 
 \end{eqnarray*}
 It follows that
\begin{eqnarray*}
\Delta 
{\rm Tr}(\mathbf{c^{\updownarrow}} \mathbf{c}^2 )&=& 
{\rm Tr}(\mathbf{x^{\updownarrow}} \mathbf{x}^2 ) - {\rm Tr}(\mathbf{c^{\updownarrow}} \mathbf{c}^2 )\nonumber
\\
&=& \sum_{ijk}\left( c^{\updownarrow}_{ij} + \Delta_{ij}^{\updownarrow} + \Delta_{ji}^{\updownarrow} \right)\left( c_{jk} + \Delta_{jk} \right)\left( c_{ki} + \Delta_{ki} \right)
 - {\rm Tr}(\mathbf{c^{\updownarrow}} \mathbf{c}^2 )
\hspace*{-10mm}
\end{eqnarray*}
Arguments similar to those employed before show that  $\sum_{j}\Delta_{ij}^{\updownarrow}\Delta_{jk} = \sum_{i}\Delta_{ij}^{\updownarrow}\Delta_{ki} = 0 $, whereas the remaining two `compound' terms give
\begin{eqnarray*}
&&\hspace*{-3mm}
\sum_{ijk}\Delta_{ji}^{\updownarrow}\Delta_{jk} c_{ki} = \\
&&~~~\sum_{ijk} 
\left( \delta_{ja}\delta_{id}c_{da} \!+\! \delta_{jc}\delta_{ib}c_{bc}\! -\! \delta_{ja}\delta_{ib}c_{ba} \!-\! \delta_{jc}\delta_{id}c_{dc}\right)
\hspace*{-10mm}
\\[-1mm]
&&~~~~~~\times \left(\delta_{ja}\delta_{kd}+ \delta_{jc}\delta_{kb} - \delta_{ja}\delta_{kb} - \delta_{jc}\delta_{kd} \right) c_{ki}\\
& &~~~= (c_{da}+c_{dc})\delta_{id}(\delta_{kd}-\delta_{kb}) c_{ki} \\
&& ~~~~~~~~~~~ + (c_{bc}+c_{ba})\delta_{ib}(\delta_{kb}-\delta_{kd}) c_{ki} \\
& &~~~= -(c_{da}+c_{dc})c_{bd} - (c_{bc}+c_{ba})c_{db} 
\end{eqnarray*}
and 
\begin{eqnarray*}
&&\hspace*{-3mm}\sum_{ijk}\Delta_{ji}^{\updownarrow} c_{jk}\Delta_{ki} =\\
&&~~~\sum_{ijk} \left( \delta_{ja}\delta_{id}c_{da} + \delta_{jc}\delta_{ib}c_{bc} - \delta_{ja}\delta_{ib}c_{ba} - \delta_{jc}\delta_{id}c_{dc}\right)
\hspace*{-10mm}
\\[-1mm]
&&~~~~~~\times\left(\delta_{ka}\delta_{id}+ \delta_{kc}\delta_{ib} - \delta_{ka}\delta_{ib} - \delta_{kc}\delta_{id} \right)c_{jk} 
\hspace*{-10mm}\\
&&~~~= (c_{da}\!+\! c_{ba})\delta_{ja}(\delta_{ka}\!-\!\delta_{kc})c_{jk} 
\\
&&~~~~~~~~~~~~+ (c_{bc}\!+\!c_{dc})\delta_{jc}(\delta_{kc}\!-\!\delta_{ka}) c_{jk}
\hspace*{-10mm} \\
&&~~~= -\left[(c_{da}+ c_{ba})c_{ac}+ (c_{bc}+c_{dc})c_{ca}\right]
\end{eqnarray*}
The product of three Deltas can be immediately seen to be zero by earlier arguments (repeated suffix in different positions). 
The other terms evaluate as follows:
\begin{eqnarray}
\sum_{ijk}\Delta_{ij}^{\updownarrow} c_{jk} c_{ki} 
&=& \sum_{\beta\in\{a,c\},~\alpha\in\{ d,b\}} \varmathbb{I}(\alpha, \beta) c_{\alpha k} c_{\alpha \beta} c_{k \beta}
\\ \nonumber
\sum_{ijk}\Delta_{ki} c_{ij}^{\updownarrow} c_{jk} 
&=& \sum_{\beta\in\{a,c\},~\alpha\in\{ d,b\}} \varmathbb{I}(\alpha, \beta) c_{\alpha k}^{\updownarrow}  c_{k \beta}
\\ \nonumber
\sum_{ijk}\Delta_{jk} c_{ij}^{\updownarrow} c_{ki}
&=& \sum_{\beta\in\{a,c\},~\alpha\in\{ d,b\}} \varmathbb{I}(\alpha, \beta) c_{\alpha k}  c_{k \beta}^{\updownarrow}
\end{eqnarray}
where $\varmathbb{I}(\alpha, \beta) $ is an indicator function which evaluates to 1 if bond $(\alpha, \beta)$ is created by the present move, to -1 if the bond $(\alpha, \beta)$ is destroyed, and zero otherwise.
Similarly
\begin{eqnarray*}
&&\hspace*{-3mm}
\sum_{ijk}\Delta_{ji}^{\updownarrow}c_{jk}c_{ki} 
=  \sum_{ijk} c_{jk}c_{ki} 
\nonumber
\\
&&~~\times\left( \delta_{ja}\delta_{id}c_{da} + \delta_{jc}\delta_{ib}c_{bc} - \delta_{ja}\delta_{ib}c_{ba} - \delta_{jc}\delta_{id}c_{dc}\right)
\nonumber
\\
&&=\sum_k\left( c_{ak}c_{kd}c_{da} \!+ \!c_{ck}c_{kb}c_{bc} \!-\! c_{ak}c_{kb}c_{ba}\! -\! c_{ck}c_{kd}c_{dc}\right)
\hspace*{-12mm}
\end{eqnarray*}
Putting all of these sub-terms together yields:
\begin{eqnarray}
&&
\hspace*{-3mm}
\Delta\left[{\rm Tr}(\mathbf{c^{\updownarrow}} \mathbf{c}^2 ) \right]
= 	\sum_{\beta\in\{a,c\},~\alpha\in\{ d,b\}} 
	\varmathbb{I}(\alpha, \beta)
	\nonumber
	\\
	&&~~~\times
	\sum_k \left[  
	 		c_{\alpha \beta}(c_{\alpha k}  c_{k \beta}+ c_{\beta k}  c_{k \alpha})
	 	+   c_{\alpha k}^{\updownarrow}  c_{k \beta}
	 	+   c_{\alpha k}  c_{k \beta}^{\updownarrow}
	 \right] \hspace*{-10mm}
\nonumber	 \\
	&& -  c_{bd} \left( c_{da} + c_{dc}\right)+ c_{ac} \left( c_{da} + c_{ba}\right) \nonumber
	\\
	&&~~~+ c_{db} \left( c_{bc} + c_{ba}\right) + c_{ca} \left( c_{bc} + c_{dc}\right)
\end{eqnarray}
\item Terms 9 and 10:
\\[1mm]
The same steps as followed to calculate term 8 can be also be applied to terms 9 and 10,  
In combination, the above ingredients lead us to the following update formula for the triangle mobility (\ref{eq:triangle_mobility_term}), as a result of the 
edge swap (\ref{square_move}):
\begin{eqnarray}
\Delta n_{\triangle} 
&=& 	\sum_{\beta \in\{1,3\},~\alpha \in\{ 4,2\}} 
	\varmathbb{I}(\alpha, \beta)
	\sum_k \left[
		c_{\alpha k} c_{k \beta}-  c_{\alpha \beta}(c_{\alpha k}  c_{k \beta}+ c_{\beta k}  c_{k \alpha})
		\right.\hspace*{-10mm}
		\nonumber
		\\
	&&\left.
	 		-   c_{\alpha k}^{\updownarrow}  c_{k \beta}
	 		-   c_{\alpha k}  c_{k \beta}^{\updownarrow}
		+  c_{\alpha k} ^{\updownarrow}  c_{k \beta}^{\updownarrow}
		-c_{\alpha \beta}
	 		\left(
	 				c_{\alpha k}^{\updownarrow}   c_{k \beta}^{\updownarrow} 
	 				+ c_{k \alpha }^{\updownarrow}   c_{\beta k }^{\updownarrow} 
	 		\right)
		\right.
		\nonumber
		\\
		&&
		\left.
		+
		 	 c_{\alpha \beta}  
	 		\left(
	 				c_{\alpha k}^{\updownarrow}  c_{k \beta}
	 				+   c_{\alpha k} c_{k \beta}^{\updownarrow} 
	 				+ c_{k \alpha }^{\updownarrow}  c_{\beta k}
	 				+   c_{k \alpha } c_{ \beta k }^{\updownarrow} 
	 		\right)
	\right]	\nonumber
\\ 
&& -
	c_{bd}(c_{db}-1)\left( c_{da}(1- c_{ba}) + c_{dc}(1-c_{bc}) \right)
			\nonumber
			\\
			&&
	 		- c_{ac} (c_{ca}-1)\left( c_{da} (1- c_{dc}) + c_{ba}(1-c_{bc} )\right)   		
	 		\nonumber 
			\\&& 
			- c_{db} ( c_{bd}-1)\left( c_{bc}(1- c_{dc} )+ c_{ba}(1-c_{da}) \right) 
	 		\nonumber 
			\\&& 
			- c_{ca} ( c_{ac}-1)\left( c_{bc}(1- c_{ba}) + c_{dc}(1-c_{da} )\right)
\end{eqnarray}
\end{itemize}

\subsection{Change in $n_{\square(\bc)}$ following one triangle-type move}

\noindent
The triangle move is a transformation from network $\bc$ to network $\mathbf{x}$, characterised by 
$x_{ij}=c_{ij} + \Omega_{ij}$ with
\begin{eqnarray}\hspace*{-3mm}
\Omega_{ij}&=& \delta_{ib}\delta_{ja}\!+\! \delta_{ic}\delta_{jb}\!+\!\delta_{ia}\delta_{jc}
\!-\!\delta_{ia}\delta_{jb}\!-\!\delta_{ib}\delta_{jc}\!-\!\delta_{ic}\delta_{ja} 
\hspace*{3mm}
\end{eqnarray}
The terms which make up the square mobility term are
\begin{eqnarray*}
 n_{\square}(\bc) &= &
 \frac{1}{2} {\rm Tr}(\bc \bc^\dag \bc \bc^\dag)
 - \sum_{ij}k^{\rm out}_i c_{ij}k^{\rm in}_j + {\rm Tr}(\bc \bc^\dag \bc)
 \nonumber
 \\&&
 + \frac{1}{2}N^2 \bra k\ket^2 + \frac{1}{2} {\rm Tr}(\bc^2)
 - \sum_i k^{\rm out}_i k^{\rm in}_i
\end{eqnarray*} 
\begin{itemize}
\item Term 2:
\begin{eqnarray*}
\sum_{ij} k^{\rm out}_i \Omega_{ij}k^{\rm in}_j = \sum_{\alpha, \beta \in \{1, 2, 3 \}} \varmathbb{I}(\alpha, \beta) k_{\alpha}^{\rm out} k_{\beta}^{\rm in}
\end{eqnarray*}
\item Term 3:
\begin{eqnarray*}
&&\hspace*{-3mm}
\Delta \Big[{\rm Tr}(\bc \bc^\dag \bc)\Big]=
{\rm Tr}(\mathbf{x} \mathbf{x}^\dag \mathbf{x})-{\rm Tr}(\bc \bc^\dag \bc)
\nonumber
\\&&= \sum_{ijk}\left( c_{ij} + \Omega_{ij}\right)\left( c_{kj} + \Omega_{kj}\right)\left( c_{ki} + \Omega_{ki}\right) - c_{ij}c_{kj}c_{ki}
\end{eqnarray*}
We consider each subterm separately:
\begin{eqnarray*}
\sum_{ijk}\Omega_{ij}c_{kj}c_{ki}&=&\sum_{ijk}\Omega_{ki}c_{kj}c_{ij}=0
\\
\sum_{ijk}\Omega_{kj}c_{ij}c_{ki} &=& \sum_{\alpha, \beta \in \{1, 2, 3 \}} \varmathbb{I}(\alpha, \beta) \sum_i c_{\alpha i} c_{i \beta}
\end{eqnarray*}
Clearly
 \begin{eqnarray*}
 \sum_{ijk}\Omega_{ij}\Omega_{kj}=\sum_{ijk}\Omega_{ki}\Omega_{kj}=0
 \end{eqnarray*}
 since the $\Omega$ kills any suffix repeated in the same position. Furthermore, 
 \begin{eqnarray*}
\sum_{i} \Omega_{ij}\Omega_{ki}=\sum_{\alpha, \beta \in \{a, b, c \}}\! \left( 1 \!-\! \delta_{\alpha \beta}\right) \delta_{j\alpha}\delta_{k \beta} - 2 \sum_{\alpha \in \{a, b, c \}} \!\delta_{j\alpha}\delta_{k \alpha}
\end{eqnarray*}
 hence 
 \begin{eqnarray*}
 \sum_{ijk}\Omega_{ij}\Omega_{ki} c_{kj} = 3
 \end{eqnarray*}
 So it follows that 
 \begin{eqnarray*}
{\rm Tr}(\mathbf{x} \mathbf{x}^\dag \mathbf{x})-{\rm Tr}(\bc \bc^\dag \bc) = 3 + \sum_{\alpha, \beta \in \{a, b, c \}} \varmathbb{I}(\alpha, \beta) 
\sum_i c_{\alpha i} c_{i \beta}
\end{eqnarray*}
\item Term 5:
\\[1mm]
We observe that $\sum_{ij}c_{ji}\Omega_{ij}=3$ and $\sum_{ij}\Omega_{ij}\Omega_{ji}=-6$. We conclude that $\Delta [ {\rm Tr}(\bc^2) ] = 0$. This is as expected, since double bonds cannot participate in a triangle swap. 
\item Term 1:
\\[1mm]
Finally we return to Term 1 using the various shortcuts derived above. We recall that a suffix repeated in the same position sends the term to zero. Hence, we already know that all terms featuring the product of 3 or 4 $\Omega$ terms will be zero. Next:
\begin{eqnarray*}
\sum_j \Omega_{ij} c_{kj} = \sum_{\alpha, \beta \in \{1, 2, 3 \}} \varmathbb{I}(\alpha, \beta) \delta_{i \alpha} c_{k \beta}
\end{eqnarray*}
From this it follows that 
\begin{eqnarray*}
\sum_{ij}\Omega_{ij} c_{kj} \Omega_{km} c_{im} = 0
\end{eqnarray*}
Finally, 
\begin{eqnarray*}
\sum_{ijkm}\Omega_{ij} c_{kj} c_{km} c_{im} &=& c_{km} \left[ c_{k1} \left( c_{2m} -c_{3m} \right) 
\right.
\\
&&\left.\hspace*{-15mm}+ c_{k2} \left( c_{3m} -c_{1m} \right)+ c_{k3} \left( c_{1m} -c_{2m} \right)\right]
\end{eqnarray*}
(and similarly with the other terms related to this one by simple permutations). 
Overall we thus find
\begin{eqnarray*}
\Delta{\rm Tr}(\bc \bc^\dag \bc \bc^\dag) 
= 4 \sum_{km}c_{km}
\sum_{\alpha, \beta \in \left\lbrace a, b, c \right\rbrace }  
\varmathbb{I}(\alpha, \beta)  
c_{\alpha m} c_{k  \beta}
\end{eqnarray*}
\end{itemize}
Collecting all these terms together, we see that the expected change in the square mobility term after the application of a single triangle type move is
\begin{eqnarray}
\Delta\left[ n_{\square}\right] = \sum_{\alpha, \beta \in \{a, b, c \}} \varmathbb{I}(\alpha, \beta) \Big[ c_{\alpha i}c_{i \beta}\!+\! k_\beta^{\rm out}k_\alpha^{\rm in} \!+\! 2 \sum_{km}c_{km}c_{\alpha m} c_{k  \beta} \Big] + 3
\nonumber
\\[-1mm]
\end{eqnarray}

\subsection{Change in $n_{\triangle}(\bc)$ following one triangle-type move}

This final incremental term is best evaluated by an algorithm which, for each edge created or destroyed, searches for mono-directed triangles that have been created or destroyed.

\end{document}